%%%%%%%%%%%%%%%%%%%%%%%%%%%%%%%%%%%%%%%%%%%%%%%%

%harvmac
%\input harvmac

%%%%%%%%%%%%%%%%%%  tex macros for preprints, cm version %%%%%%%%%%%%%%
%                     (P. Ginsparg, last updated 9/91)
%                if confused, type `b' in response to query
%
%---------------------------------------------------------------------%
%% site dependent options:
%% \unredoffs and \redoffs define horizontal and vertical offsets
%% respectively for unreduced and reduced modes. \speclscape defines
%% the \special{} call that sets printer to landscape (sideways) mode.
%% from standard set below, leave uncommented as appropriate or redefine
%
%%% next 400dpi
%\def\unredoffs{} \def\redoffs{\voffset=-.31truein\hoffset=-.48truein}
%\def\speclscape{\special{landscape}}
%
%%% apple lw
\def\unredoffs{} \def\redoffs{\voffset=-.31truein\hoffset=-.59truein}
\def\speclscape{\special{ps: landscape}}
%
%%% qms lasergrafix:
%\def\unredoffs{} \def\redoffs{\voffset=-.4truein\hoffset=.125truein}
%\def\speclscape{\special{qms: landscape}}
%
%%% saclay A4 paper:
%\def\unredoffs{\hoffset-.14truein\voffset-.2truein}
%\def\redoffs{\voffset=-.45truein\hoffset=-.21truein}
%\def\speclscape{\special{landscape}}
%
%---------------------------------------------------------------------%
%
\newbox\leftpage \newdimen\fullhsize \newdimen\hstitle \newdimen\hsbody
\tolerance=1000\hfuzz=2pt
\catcode`\@=11 % This allows us to modify PLAIN macros.
\def\bigans{b }
\def\answ{b }

%\message{ big or little (b/l)? }\read-1 to\answ
%
\ifx\answ\bigans\message{(This will come out unreduced.}
\magnification=1200\unredoffs\baselineskip=16pt plus 2pt minus 1pt
\hsbody=\hsize \hstitle=\hsize %take default values for unreduced format
\else\message{(This will be reduced.} \let\l@r=L
\magnification=1000\baselineskip=16pt plus 2pt minus 1pt \vsize=7truein
\redoffs \hstitle=8truein\hsbody=4.75truein\fullhsize=10truein\hsize=\hsbody
\output={\ifnum\pageno=0 %%% This is the HUTP version
  \shipout\vbox{\speclscape{\hsize\fullhsize\makeheadline}
    \hbox to \fullhsize{\hfill\pagebody\hfill}}\advancepageno
  \else
  \almostshipout{\leftline{\vbox{\pagebody\makefootline}}}\advancepageno
  \fi}
\def\almostshipout#1{\if L\l@r \count1=1 \message{[\the\count0.\the\count1]}
      \global\setbox\leftpage=#1 \global\let\l@r=R
 \else \count1=2
  \shipout\vbox{\speclscape{\hsize\fullhsize\makeheadline}
      \hbox to\fullhsize{\box\leftpage\hfil#1}}  \global\let\l@r=L\fi}
\fi
%---------------------------------------------------------------------
%
\newcount\yearltd\yearltd=\year\advance\yearltd by -1900

\def\Title#1#2{\nopagenumbers\abstractfont\hsize=\hstitle\rightline{#1}%
\vskip 1in\centerline{\titlefont #2}\abstractfont\vskip .5in\pageno=0}
\def\Date#1{\vfill\leftline{#1}\tenpoint\supereject\global\hsize=\hsbody%
\footline={\hss\tenrm\folio\hss}}%  restores pagenumbers
%
%       use following instead of \Date on the preliminary draft,
%       puts date/time on each page in big mode, writes labels in margins

\def\draftmode{\message{ DRAFTMODE }\def\draftdate{{\rm preliminary draft:
\number\month/\number\day/\number\yearltd\ \ \hourmin}}%
\headline={\hfil\draftdate}\writelabels\baselineskip=20pt plus 2pt minus 2pt
 {\count255=\time\divide\count255 by 60 \xdef\hourmin{\number\count255}
  \multiply\count255 by-60\advance\count255 by\time
  \xdef\hourmin{\hourmin:\ifnum\count255<10 0\fi\the\count255}}}
%       use \nolabels to get rid of eqn, ref, and fig labels in draft mode
\def\nolabels{\def\wrlabeL##1{}\def\eqlabeL##1{}\def\reflabeL##1{}}
\def\writelabels{\def\wrlabeL##1{\leavevmode\vadjust{\rlap{\smash%
{\line{{\escapechar=` \hfill\rlap{\sevenrm\hskip.03in\string##1}}}}}}}%
\def\eqlabeL##1{{\escapechar-1\rlap{\sevenrm\hskip.05in\string##1}}}%
\def\reflabeL##1{\noexpand\llap{\noexpand\sevenrm\string\string\string##1}}}
\nolabels
%
% tagged sec numbers
\global\newcount\secno \global\secno=0
\global\newcount\meqno \global\meqno=1
\def\newsec#1{\global\advance\secno by1\message{(\the\secno. #1)}
%\ifx\answ\bigans \vfill\eject \else \bigbreak\bigskip \fi  %if desired
\global\subsecno=0\eqnres@t\noindent{\bf\the\secno. #1}
\writetoca{{\secsym} {#1}}\par\nobreak\medskip\nobreak}
\def\eqnres@t{\xdef\secsym{\the\secno.}\global\meqno=1\bigbreak\bigskip}
\def\sequentialequations{\def\eqnres@t{\bigbreak}}\xdef\secsym{}
\global\newcount\subsecno \global\subsecno=0
\def\subsec#1{\global\advance\subsecno by1\message{(\secsym\the\subsecno. #1)}
\ifnum\lastpenalty>9000\else\bigbreak\fi
\noindent{\it\secsym\the\subsecno. #1}\writetoca{\string\quad
{\secsym\the\subsecno.} {#1}}\par\nobreak\medskip\nobreak}
\def\appendix#1#2{\global\meqno=1\global\subsecno=0\xdef\secsym{\hbox{#1.}}
\bigbreak\bigskip\noindent{\bf Appendix #1. #2}\message{(#1. #2)}
\writetoca{Appendix {#1.} {#2}}\par\nobreak\medskip\nobreak}
%
%       \eqn\label{a+b=c}   gives displayed equation, numbered
%               consecutively within sections.
%     \eqnn and \eqna define labels in advance (of eqalign?)
%
\def\eqnn#1{\xdef #1{(\secsym\the\meqno)}\writedef{#1\leftbracket#1}%
\global\advance\meqno by1\wrlabeL#1}
\def\eqna#1{\xdef #1##1{\hbox{$(\secsym\the\meqno##1)$}}
\writedef{#1\numbersign1\leftbracket#1{\numbersign1}}%
\global\advance\meqno by1\wrlabeL{#1$\{\}$}}
\def\eqn#1#2{\xdef #1{(\secsym\the\meqno)}\writedef{#1\leftbracket#1}%
\global\advance\meqno by1$$#2\eqno#1\eqlabeL#1$$}
%
%            footnotes
\newskip\footskip\footskip14pt plus 1pt minus 1pt %sets footnote baselineskip
\def\footnotefont{\ninepoint}\def\f@t#1{\footnotefont #1\@foot}
\def\f@@t{\baselineskip\footskip\bgroup\footnotefont\aftergroup\@foot\let\next}
\setbox\strutbox=\hbox{\vrule height9.5pt depth4.5pt width0pt}
\global\newcount\ftno \global\ftno=0
\def\foot{\global\advance\ftno by1\footnote{$^{\the\ftno}$}}
%
%say \footend to put footnotes at end
%will cause problems if \ref used inside \foot, instead use \nref before
\newwrite\ftfile
\def\footend{\def\foot{\global\advance\ftno by1\chardef\wfile=\ftfile
$^{\the\ftno}$\ifnum\ftno=1\immediate\openout\ftfile=foots.tmp\fi%
\immediate\write\ftfile{\noexpand\smallskip%
\noexpand\item{f\the\ftno:\ }\pctsign}\findarg}%
\def\footatend{\vfill\eject\immediate\closeout\ftfile{\parindent=20pt
\centerline{\bf Footnotes}\nobreak\bigskip\input foots.tmp }}}
\def\footatend{}
%
%     \ref\label{text}
% generates a number, assigns it to \label, generates an entry.
% To list the refs on a separate page,  \listrefs
%
\global\newcount\refno \global\refno=1
\newwrite\rfile
\def\ref{[\the\refno]\nref}
\def\nref#1{\xdef#1{[\the\refno]}\writedef{#1\leftbracket#1}%
\ifnum\refno=1\immediate\openout\rfile=refs.tmp\fi
\global\advance\refno by1\chardef\wfile=\rfile\immediate
\write\rfile{\noexpand\item{#1\ }\reflabeL{#1\hskip.31in}\pctsign}\findarg}
%   horrible hack to sidestep tex \write limitation
\def\findarg#1#{\begingroup\obeylines\newlinechar=`\^^M\pass@rg}
{\obeylines\gdef\pass@rg#1{\writ@line\relax #1^^M\hbox{}^^M}%
\gdef\writ@line#1^^M{\expandafter\toks0\expandafter{\striprel@x #1}%
\edef\next{\the\toks0}\ifx\next\em@rk\let\next=\endgroup\else\ifx\next\empty%
\else\immediate\write\wfile{\the\toks0}\fi\let\next=\writ@line\fi\next\relax}}
\def\striprel@x#1{} \def\em@rk{\hbox{}}
\def\lref{\begingroup\obeylines\lr@f}
\def\lr@f#1#2{\gdef#1{\ref#1{#2}}\endgroup\unskip}

\def\addref#1{\immediate\write\rfile{\noexpand\item{}#1}} %now unnecessary
\def\startrefs#1{\immediate\openout\rfile=refs.tmp\refno=#1}
\def\xref{\expandafter\xr@f}\def\xr@f[#1]{#1}
\def\refs#1{\count255=1[\r@fs #1{\hbox{}}]}
\def\r@fs#1{\ifx\und@fined#1\message{reflabel \string#1 is undefined.}%
\nref#1{need to supply reference \string#1.}\fi%
\vphantom{\hphantom{#1}}\edef\next{#1}\ifx\next\em@rk\def\next{}%
\else\ifx\next#1\ifodd\count255\relax\xref#1\count255=0\fi%
\else#1\count255=1\fi\let\next=\r@fs\fi\next}
%

%
% this is ugly, but moore insists
\newwrite\ffile\global\newcount\figno \global\figno=1
\def\fig{fig.~\the\figno\nfig}
\def\nfig#1{\xdef#1{fig.~\the\figno}%
\writedef{#1\leftbracket fig.\noexpand~\the\figno}%
\ifnum\figno=1\immediate\openout\ffile=figs.tmp\fi\chardef\wfile=\ffile%
\immediate\write\ffile{\noexpand\medskip\noexpand\item{Fig.\ \the\figno. }
\reflabeL{#1\hskip.55in}\pctsign}\global\advance\figno by1\findarg}
\def\xfig{\expandafter\xf@g}\def\xf@g fig.\penalty\@M\ {}
\def\figs#1{figs.~\f@gs #1{\hbox{}}}
\def\f@gs#1{\edef\next{#1}\ifx\next\em@rk\def\next{}\else
\ifx\next#1\xfig #1\else#1\fi\let\next=\f@gs\fi\next}
\newwrite\lfile
{\escapechar-1\xdef\pctsign{\string\%}\xdef\leftbracket{\string\{}
\xdef\rightbracket{\string\}}\xdef\numbersign{\string\#}}

\def\writestop{\def\writestoppt{\immediate\write\lfile{\string\pageno%
\the\pageno\string\startrefs\leftbracket\the\refno\rightbracket%
\string\def\string\secsym\leftbracket\secsym\rightbracket%
\string\secno\the\secno\string\meqno\the\meqno}\immediate\closeout\lfile}}
\def\writestoppt{}\def\writedef#1{}
\def\seclab#1{\xdef #1{\the\secno}\writedef{#1\leftbracket#1}\wrlabeL{#1=#1}}
\def\subseclab#1{\xdef #1{\secsym\the\subsecno}%
\writedef{#1\leftbracket#1}\wrlabeL{#1=#1}}
\newwrite\tfile \def\writetoca#1{}
\def\leaderfill{\leaders\hbox to 1em{\hss.\hss}\hfill}
%   use this to write file with table of contents
\def\writetoc{\immediate\openout\tfile=toc.tmp
   \def\writetoca##1{{\edef\next{\write\tfile{\noindent ##1
   \string\leaderfill {\noexpand\number\pageno} \par}}\next}}}
%       and this lists table of contents on second pass

%
\catcode`\@=12 % at signs are no longer letters
%
%   Unpleasantness in calling in abstract and title fonts
\edef\tfontsize{\ifx\answ\bigans scaled\magstep3\else scaled\magstep4\fi}
\font\titlerm=cmr10 \tfontsize \font\titlerms=cmr7 \tfontsize
\font\titlermss=cmr5 \tfontsize \font\titlei=cmmi10 \tfontsize
\font\titleis=cmmi7 \tfontsize \font\titleiss=cmmi5 \tfontsize
\font\titlesy=cmsy10 \tfontsize \font\titlesys=cmsy7 \tfontsize
\font\titlesyss=cmsy5 \tfontsize \font\titleit=cmti10 \tfontsize
\skewchar\titlei='177 \skewchar\titleis='177 \skewchar\titleiss='177
\skewchar\titlesy='60 \skewchar\titlesys='60 \skewchar\titlesyss='60
\def\titlefont{\def\rm{\fam0\titlerm}% switch to title font
\textfont0=\titlerm \scriptfont0=\titlerms \scriptscriptfont0=\titlermss
\textfont1=\titlei \scriptfont1=\titleis \scriptscriptfont1=\titleiss
\textfont2=\titlesy \scriptfont2=\titlesys \scriptscriptfont2=\titlesyss
\textfont\itfam=\titleit \def\it{\fam\itfam\titleit}\rm}
 \ifx\answ\bigans\else scaled\magstep1\fi
\ifx\answ\bigans\def\abstractfont{\tenpoint}\else
\font\abssl=cmsl10 scaled \magstep1
\font\absrm=cmr10 scaled\magstep1 \font\absrms=cmr7 scaled\magstep1
\font\absrmss=cmr5 scaled\magstep1 \font\absi=cmmi10 scaled\magstep1
\font\absis=cmmi7 scaled\magstep1 \font\absiss=cmmi5 scaled\magstep1
\font\abssy=cmsy10 scaled\magstep1 \font\abssys=cmsy7 scaled\magstep1
\font\abssyss=cmsy5 scaled\magstep1 \font\absbf=cmbx10 scaled\magstep1
\skewchar\absi='177 \skewchar\absis='177 \skewchar\absiss='177
\skewchar\abssy='60 \skewchar\abssys='60 \skewchar\abssyss='60
\def\abstractfont{\def\rm{\fam0\absrm}% switch to abstract font
\textfont0=\absrm \scriptfont0=\absrms \scriptscriptfont0=\absrmss
\textfont1=\absi \scriptfont1=\absis \scriptscriptfont1=\absiss
\textfont2=\abssy \scriptfont2=\abssys \scriptscriptfont2=\abssyss
\textfont\itfam=\bigit \def\it{\fam\itfam\bigit}\def\footnotefont{\tenpoint}%
\textfont\slfam=\abssl \def\sl{\fam\slfam\abssl}%
\textfont\bffam=\absbf \def\bf{\fam\bffam\absbf}\rm}\fi
\def\tenpoint{\def\rm{\fam0\tenrm}% switch back to 10-point type
\textfont0=\tenrm \scriptfont0=\sevenrm \scriptscriptfont0=\fiverm
\textfont1=\teni  \scriptfont1=\seveni  \scriptscriptfont1=\fivei
\textfont2=\tensy \scriptfont2=\sevensy \scriptscriptfont2=\fivesy
\textfont\itfam=\tenit \def\it{\fam\itfam\tenit}\def\footnotefont{\ninepoint}%
\textfont\bffam=\tenbf \def\bf{\fam\bffam\tenbf}\def\sl{\fam\slfam\tensl}\rm}
\font\ninerm=cmr9 \font\sixrm=cmr6 \font\ninei=cmmi9 \font\sixi=cmmi6
\font\ninesy=cmsy9 \font\sixsy=cmsy6 \font\ninebf=cmbx9
\font\nineit=cmti9 \font\ninesl=cmsl9 \skewchar\ninei='177
\skewchar\sixi='177 \skewchar\ninesy='60 \skewchar\sixsy='60
\def\ninepoint{\def\rm{\fam0\ninerm}% switch to footnote font
\textfont0=\ninerm \scriptfont0=\sixrm \scriptscriptfont0=\fiverm
\textfont1=\ninei \scriptfont1=\sixi \scriptscriptfont1=\fivei
\textfont2=\ninesy \scriptfont2=\sixsy \scriptscriptfont2=\fivesy
\textfont\itfam=\ninei \def\it{\fam\itfam\nineit}\def\sl{\fam\slfam\ninesl}%
\textfont\bffam=\ninebf \def\bf{\fam\bffam\ninebf}\rm}
%
%---------------------------------------------------------------------
%

\hyphenation{anom-aly anom-alies coun-ter-term coun-ter-terms}
\def\inv{^{\raise.15ex\hbox{${\scriptscriptstyle -}$}\kern-.05em 1}}

\def\Dsl{\,\raise.15ex\hbox{/}\mkern-13.5mu D} %this one can be subscripted
\def\dsl{\raise.15ex\hbox{/}\kern-.57em\partial}

\font\bigit=cmti10 scaled \magstep1
 %pound sterling
\def\lspace{\ifx\answ\bigans{}\else\qquad\fi}
\def\lbspace{\ifx\answ\bigans{}\else\hskip-.2in\fi} % $$\lbspace...$$
\def\boxeqn#1{\vcenter{\vbox{\hrule\hbox{\vrule\kern3pt\vbox{\kern3pt
    \hbox{${\displaystyle #1}$}\kern3pt}\kern3pt\vrule}\hrule}}}
\def\mbox#1#2{\vcenter{\hrule \hbox{\vrule height#2in
        \kern#1in \vrule} \hrule}}  %e.g. \mbox{.1}{.1}
%   matters of taste
%\def\tilde{\widetilde} \def\bar{\overline} \def\hat{\widehat}
%
% some sample definitions
  %     curly letters

\def\darr#1{\raise1.5ex\hbox{$\leftrightarrow$}\mkern-16.5mu #1}
 %pound sterling

 %puts a small half in a displayed eqn
\def\roughly#1{\raise.3ex\hbox{$#1$\kern-.75em\lower1ex\hbox{$\sim$}}}

%\draftmode
\let\includefigures=\iftrue
\let\useblackboard=\iftrue
\newfam\black

%Figure Stuff
\includefigures
\message{If you do not have epsf.tex (to include figures),}
\message{change the option at the top of the tex file.}
\input epsf
\def\figin{\epsfcheck\figin}\def\figins{\epsfcheck\figins}
\def\epsfcheck{\ifx\epsfbox\UnDeFiNeD
\message{(NO epsf.tex, FIGURES WILL BE IGNORED)}
\gdef\figin##1{\vskip2in}\gdef\figins##1{\hskip.5in}% blank space instead
\else\message{(FIGURES WILL BE INCLUDED)}%
\gdef\figin##1{##1}\gdef\figins##1{##1}\fi}
\def\DefWarn#1{}
\def\figinsert{\goodbreak\midinsert}
\def\ifig#1#2#3{\DefWarn#1\xdef#1{fig.~\the\figno}
\writedef{#1\leftbracket fig.\noexpand~\the\figno}%
\figinsert\figin{\centerline{#3}}\medskip\centerline{\vbox{
\baselineskip12pt\advance\hsize by -1truein
\noindent\footnotefont{\bf Fig.~\the\figno:} #2}}
%\bigskip
\endinsert\global\advance\figno by1}
%%%
\else
\def\ifig#1#2#3{\xdef#1{fig.~\the\figno}
\writedef{#1\leftbracket fig.\noexpand~\the\figno}%
%\figinsert\figin{\centerline{#3}}\medskip
%\centerline{\vbox{\baselineskip12pt
%\advance\hsize by -1truein\noindent
%\footnotefont{\bf Fig.~\the\figno:} #2}}
%\bigskip\endinsert
\global\advance\figno by1} \fi

\def\id{{1 \kern-.28em {\rm l}}}

\def\K3{{\bf K3}}
\def\journal#1&#2(#3){\unskip, \sl #1\ \bf #2 \rm(19#3) }
\def\andjournal#1&#2(#3){\sl #1~\bf #2 \rm (19#3) }

\def\ie{{\it i.e.}}

\def\tilde{\widetilde}

\def\frac#1#2{{#1\over#2}}

\def\inbar{\,\vrule height1.5ex width.4pt depth0pt}
\def\IC{\relax\hbox{$\inbar\kern-.3em{\rm C}$}}
\def\IR{\relax{\rm I\kern-.18em R}}
\def\IP{\relax{\rm I\kern-.18em P}}

%
%%%%%%%%%%%%%%%%%%%%%%%%%%%%%%%%%%%%
%

%
\catcode`\@=11
\def\slash#1{\mathord{\mathpalette\c@ncel{#1}}}
\overfullrule=0pt

\def\LL{{\cal L}}
\def\NN{{\cal N}}
\def\OO{{\cal O}}

\def\underrel#1\over#2{\mathrel{\mathop{\kern\z@#1}\limits_{#2}}}

\catcode`\@=12

%%%%%%%%%%%%%%%%%%%%%%%%%%%%%%%%%%%%%%%%%%%%%%%%%%%%%%%%%%%%%%

%

\def\det{{\rm det}}

\def\det{{\rm det}}
\def\exp{{\rm exp}}

%%%%%%%%%%%%%%%%%%%%%%%%%%%%%%%%%%%%%%%%%%%%%%%%%%%%%%%%%%%%%%
% new defs:

\def\p{{\partial}}

\def\ra{{\rightarrow}}

\def\LL{{\cal L}}

%\SachdevDQ
\lref\SachdevDQ{
  S.~Sachdev,
  ``The Quantum phases of matter,''
[arXiv:1203.4565 [hep-th]].
%%CITATION = arXiv:1203.4565%%
}

%\LiuDM
\lref\LiuDM{
  H.~Liu, J.~McGreevy and D.~Vegh,
  ``Non-Fermi liquids from holography,''
Phys.\ Rev.\ D {\bf 83}, 065029 (2011).
[arXiv:0903.2477 [hep-th]].
%%CITATION = arXiv:0903.2477%%
}

%\ShaghoulianAA
\lref\ShaghoulianAA{
  E.~Shaghoulian,
  ``Holographic Entanglement Entropy and Fermi Surfaces,''
JHEP {\bf 1205}, 065 (2012).
[arXiv:1112.2702 [hep-th]].
%%CITATION = arXiv:1112.2702%%
}

%\DongSE
\lref\DongSE{
  X.~Dong, S.~Harrison, S.~Kachru, G.~Torroba and H.~Wang,
  ``Aspects of holography for theories with hyperscaling violation,''
JHEP {\bf 1206}, 041 (2012).
[arXiv:1201.1905 [hep-th]].
%%CITATION = arXiv:1201.1905%%
}

%\NarayanHK
\lref\NarayanHK{
  K.~Narayan,
  ``On Lifshitz scaling and hyperscaling violation in string theory,''
Phys.\ Rev.\ D {\bf 85}, 106006 (2012).
[arXiv:1202.5935 [hep-th]].
%%CITATION = arXiv:1202.5935%%
}

%\KimNB
\lref\KimNB{
  B.~S.~Kim,
  ``Schr\'odinger Holography with and without Hyperscaling Violation,''
JHEP {\bf 1206}, 116 (2012).
[arXiv:1202.6062 [hep-th]].
%%CITATION = arXiv:1202.6062%%
}

%\HashimotoTI
\lref\HashimotoTI{
  K.~Hashimoto and N.~Iizuka,
  ``A Comment on Holographic Luttinger Theorem,''
[arXiv:1203.5388 [hep-th]].
%%CITATION = arXiv:1203.5388%%
}

%\PerlmutterHE
\lref\PerlmutterHE{
  E.~Perlmutter,
  ``Hyperscaling violation from supergravity,''
JHEP {\bf 1206}, 165 (2012).
[arXiv:1205.0242 [hep-th]].
%%CITATION = arXiv:1205.0242%%
}

%\CubrovicYE
\lref\CubrovicYE{
  M.~Cubrovic, J.~Zaanen and K.~Schalm,
  ``String Theory, Quantum Phase Transitions and the Emergent Fermi-Liquid,''
Science {\bf 325}, 439 (2009).
[arXiv:0904.1993 [hep-th]].
%%CITATION = arXiv:0904.1993%%
}

%\FaulknerWJ
\lref\FaulknerWJ{
  T.~Faulkner, H.~Liu, J.~McGreevy and D.~Vegh,
  ``Emergent quantum criticality, Fermi surfaces, and AdS(2),''
Phys.\ Rev.\ D {\bf 83}, 125002 (2011).
[arXiv:0907.2694 [hep-th]].
%%CITATION = arXiv:0907.2694%%
}

%\GubserQT
\lref\GubserQT{
  S.~S.~Gubser and F.~D.~Rocha,
  ``Peculiar properties of a charged dilatonic black hole in $AdS_5$,''
Phys.\ Rev.\ D {\bf 81}, 046001 (2010).
[arXiv:0911.2898 [hep-th]].
%%CITATION = arXiv:0911.2898%%
}

%\GubserYB
\lref\GubserYB{
  S.~S.~Gubser and J.~Ren,
  ``Analytic fermionic Green's functions from holography,''
[arXiv:1204.6315 [hep-th]].
%%CITATION = arXiv:1204.6315%%
}

%\HuijseEF
\lref\HuijseEF{
  L.~Huijse, S.~Sachdev and B.~Swingle,
  ``Hidden Fermi surfaces in compressible states of gauge-gravity duality,''
Phys.\ Rev.\ B {\bf 85}, 035121 (2012).
[arXiv:1112.0573 [cond-mat.str-el]].
%%CITATION = arXiv:1112.0573%%
}

%\OgawaBZ
\lref\OgawaBZ{
  N.~Ogawa, T.~Takayanagi and T.~Ugajin,
  ``Holographic Fermi Surfaces and Entanglement Entropy,''
JHEP {\bf 1201}, 125 (2012).
[arXiv:1111.1023 [hep-th]].
%%CITATION = arXiv:1111.1023%%
}

%\HartnollWM
\lref\HartnollWM{
  S.~A.~Hartnoll and E.~Shaghoulian,
  ``Spectral weight in holographic scaling geometries,''
[arXiv:1203.4236 [hep-th]].
%%CITATION = arXiv:1203.4236%%
}

%\RyuBV
\lref\RyuBV{
  S.~Ryu and T.~Takayanagi,
  ``Holographic derivation of entanglement entropy from AdS/CFT,''
Phys.\ Rev.\ Lett.\  {\bf 96}, 181602 (2006).
[hep-th/0603001].
%%CITATION = hep-th/0603001%%
}

%\RyuEF
\lref\RyuEF{
  S.~Ryu and T.~Takayanagi,
  ``Aspects of Holographic Entanglement Entropy,''
JHEP {\bf 0608}, 045 (2006).
[hep-th/0605073].
%%CITATION = hep-th/0605073%%
}

%\SwingleBF
\lref\SwingleBF{
  B.~Swingle,
  ``Entanglement Entropy and the Fermi Surface,''
[arXiv:0908.1724 [cond-mat.str-el]].
%%CITATION = arXiv:0908.1724%%
}

%\ZhangRW
\lref\ZhangRW{
  Y.~Zhang, T.~Grover and A.~Vishwanath,
  ``Topological Entanglement Entropy of Z2 Spin liquids and Lattice Laughlin states,''
Phys.\ Rev.\ B {\bf 84}, 075128 (2011).
[arXiv:1106.0015 [cond-mat.str-el]].
%%CITATION = arXiv:1106.0015%%
}

\lref\DSY{
W. Ding, A. Seidel, and K. Yang, 
   "Entanglement Entropy of Fermi Liquids via Multi-dimensional Bosonization," 
   [arXiv:1110.3004 [cond-mat.stat-mech]].
  }
  
\lref\WGK{
M.M. Wolf, 
"Violation of the entropic area law for fermions," 
Phys. Rev. Lett. 96 (2006) 010404;
D. Gioev and I. Klich, 
"Entanglement entropy of fermions in any dimension and the
Widom conjecture," 
Phys. Rev. Lett. 96 (2006) 100503.
}

%\LiuEE
\lref\LiuEE{
  H.~Liu and M.~Mezei,
  ``A Refinement of entanglement entropy and the number of degrees of freedom,''
[arXiv:1202.2070 [hep-th]].
%%CITATION = arXiv:1202.2070%%
}

%\PakmanUI
\lref\PakmanUI{
  A.~Pakman and A.~Parnachev,
  ``Topological Entanglement Entropy and Holography,''
JHEP {\bf 0807}, 097 (2008).
[arXiv:0805.1891 [hep-th]].
%%CITATION = arXiv:0805.1891%%
}

%\TaylorTG
\lref\TaylorTG{
  M.~Taylor,
  ``Non-relativistic holography,''
[arXiv:0812.0530 [hep-th]].
%%CITATION = arXiv:0812.0530%%
}

%\GoldsteinAW
\lref\GoldsteinAW{
  K.~Goldstein, N.~Iizuka, S.~Kachru, S.~Prakash, S.~P.~Trivedi and A.~Westphal,
  ``Holography of Dyonic Dilaton Black Branes,''
JHEP {\bf 1010}, 027 (2010).
[arXiv:1007.2490 [hep-th]].
%%CITATION = arXiv:1007.2490%%
}

%\CharmousisZZ
\lref\CharmousisZZ{
  C.~Charmousis, B.~Gouteraux, B.~S.~Kim, E.~Kiritsis and R.~Meyer,
  ``Effective Holographic Theories for low-temperature condensed matter systems,''
JHEP {\bf 1011}, 151 (2010).
[arXiv:1005.4690 [hep-th]].
%%CITATION = arXiv:1005.4690%%
}

%\CadoniKV
\lref\CadoniKV{
  M.~Cadoni and P.~Pani,
  ``Holography of charged dilatonic black branes at finite temperature,''
JHEP {\bf 1104}, 049 (2011).
[arXiv:1102.3820 [hep-th]].
%%CITATION = arXiv:1102.3820%%
}

%\GouterauxCE
\lref\GouterauxCE{
  B.~Gouteraux and E.~Kiritsis,
  ``Generalized Holographic Quantum Criticality at Finite Density,''
JHEP {\bf 1112}, 036 (2011).
[arXiv:1107.2116 [hep-th]].
%%CITATION = arXiv:1107.2116%%
}

%\KlebanovWS
\lref\KlebanovWS{
  I.~R.~Klebanov, D.~Kutasov and A.~Murugan,
  ``Entanglement as a probe of confinement,''
Nucl.\ Phys.\ B {\bf 796}, 274 (2008).
[arXiv:0709.2140 [hep-th]].
%%CITATION = arXiv:0709.2140%%
}

%\DongSE
\lref\DongSE{
  X.~Dong, S.~Harrison, S.~Kachru, G.~Torroba and H.~Wang,
  ``Aspects of holography for theories with hyperscaling violation,''
JHEP {\bf 1206}, 041 (2012).
[arXiv:1201.1905 [hep-th]].
%%CITATION = arXiv:1201.1905%%
}

%\CveticXP
\lref\CveticXP{
  M.~Cvetic, M.~J.~Duff, P.~Hoxha, J.~T.~Liu, H.~Lu, J.~X.~Lu, R.~Martinez-Acosta and C.~N.~Pope {\it et al.},
  ``Embedding AdS black holes in ten-dimensions and eleven-dimensions,''
Nucl.\ Phys.\ B {\bf 558}, 96 (1999).
[hep-th/9903214].
%%CITATION = hep-th/9903214%%
}

%\FursaevIH
\lref\FursaevIH{
  D.~V.~Fursaev,
  ``Proof of the holographic formula for entanglement entropy,''
JHEP {\bf 0609}, 018 (2006).
[hep-th/0606184].
%%CITATION = hep-th/0606184%%
}

%\deBoerWK
\lref\deBoerWK{
  J.~de Boer, M.~Kulaxizi and A.~Parnachev,
  ``Holographic Entanglement Entropy in Lovelock Gravities,''
JHEP {\bf 1107}, 109 (2011).
[arXiv:1101.5781 [hep-th]].
%%CITATION = arXiv:1101.5781%%
}

%\HungXB
\lref\HungXB{
  L.~-Y.~Hung, R.~C.~Myers and M.~Smolkin,
  ``On Holographic Entanglement Entropy and Higher Curvature Gravity,''
JHEP {\bf 1104}, 025 (2011).
[arXiv:1101.5813 [hep-th]].
%%CITATION = arXiv:1101.5813%%
}

%\BriganteNU
\lref\BriganteNU{
  M.~Brigante, H.~Liu, R.~C.~Myers, S.~Shenker and S.~Yaida,
  ``Viscosity Bound Violation in Higher Derivative Gravity,''
Phys.\ Rev.\ D {\bf 77}, 126006 (2008).
[arXiv:0712.0805 [hep-th]].
%%CITATION = arXiv:0712.0805%%
}

%\BriganteGZ
\lref\BriganteGZ{
  M.~Brigante, H.~Liu, R.~C.~Myers, S.~Shenker and S.~Yaida,
  %\``The Viscosity Bound and Causality Violation,''
Phys.\ Rev.\ Lett.\  {\bf 100}, 191601 (2008).
[arXiv:0802.3318 [hep-th]].
%%CITATION = arXiv:0802.3318%%
}

%\BuchelTT
\lref\BuchelTT{
  A.~Buchel and R.~C.~Myers,
  ``Causality of Holographic Hydrodynamics,''
JHEP {\bf 0908}, 016 (2009).
[arXiv:0906.2922 [hep-th]].
%%CITATION = arXiv:0906.2922%%
}

%\HofmanUG
\lref\HofmanUG{
  D.~M.~Hofman,
  ``Higher Derivative Gravity, Causality and Positivity of Energy in a UV complete QFT,''
Nucl.\ Phys.\ B {\bf 823}, 174 (2009).
[arXiv:0907.1625 [hep-th]].
%%CITATION = arXiv:0907.1625%%
}

%\AmmonJE
\lref\AmmonJE{
  M.~Ammon, M.~Kaminski and A.~Karch,
  ``Hyperscaling-Violation on Probe D-Branes,''
[arXiv:1207.1726 [hep-th]].
%%CITATION = arXiv:1207.1726%%
}

%\DeWolfeUV
\lref\DeWolfeUV{
  O.~DeWolfe, S.~S.~Gubser and C.~Rosen,
  ``Fermi surfaces in N=4 Super-Yang-Mills theory,''
[arXiv:1207.3352 [hep-th]].
%%CITATION = arXiv:1207.3352%%
}

%\CadoniUF
\lref\CadoniUF{
  M.~Cadoni and S.~Mignemi,
  ``Phase transition and hyperscaling violation for scalar Black Branes,''
JHEP {\bf 1206}, 056 (2012).
[arXiv:1205.0412 [hep-th]].
%%CITATION = arXiv:1205.0412%%
}

%\NishiokaGR
\lref\NishiokaGR{
  T.~Nishioka and T.~Takayanagi,
  ``AdS Bubbles, Entropy and Closed String Tachyons,''
JHEP {\bf 0701}, 090 (2007).
[hep-th/0611035].
%%CITATION = hep-th/0611035%%
}

%\KlebanovYF
\lref\KlebanovYF{
  I.~R.~Klebanov, T.~Nishioka, S.~S.~Pufu and B.~R.~Safdi,
  ``On Shape Dependence and RG Flow of Entanglement Entropy,''
JHEP {\bf 1207}, 001 (2012).
[arXiv:1204.4160 [hep-th]].
%%CITATION = arXiv:1204.4160%%
}

%\CasiniSR
\lref\CasiniSR{
  H.~Casini and M.~Huerta,
  ``Entanglement entropy in free quantum field theory,''
J.\ Phys.\ A A {\bf 42}, 504007 (2009).
[arXiv:0905.2562 [hep-th]].
%%CITATION = arXiv:0905.2562%%
}

%\GrossGK
\lref\GrossGK{
  D.~J.~Gross and H.~Ooguri,
  ``Aspects of large N gauge theory dynamics as seen by string theory,''
Phys.\ Rev.\ D {\bf 58}, 106002 (1998).
[hep-th/9805129].
%%CITATION = hep-th/9805129%%
}

%\OlesenJI
\lref\OlesenJI{
  P.~Olesen and K.~Zarembo,
  ``Phase transition in Wilson loop correlator from AdS / CFT correspondence,''
[hep-th/0009210].
%%CITATION = hep-th/0009210%%
}

\lref\ZaremboBU{
  K.~Zarembo,
  ``Wilson loop correlator in the AdS / CFT correspondence,''
Phys.\ Lett.\ B {\bf 459}, 527 (1999).
[hep-th/9904149].
%%CITATION = hep-th/9904149%%
}

\lref\KimTD{
  H.~Kim, D.~K.~Park, S.~Tamarian and H.~J.~W.~Muller-Kirsten,
  ``Gross-Ooguri phase transition at zero and finite temperature: Two circular Wilson loop case,''
JHEP {\bf 0103}, 003 (2001).
[hep-th/0101235].
%%CITATION = hep-th/0101235%%
}

%\HirataJX
\lref\HirataJX{
  T.~Hirata and T.~Takayanagi,
  ``AdS/CFT and strong subadditivity of entanglement entropy,''
JHEP {\bf 0702}, 042 (2007).
[hep-th/0608213].
%%CITATION = hep-th/0608213%%
}

%\AlishahihaAD
\lref\AlishahihaAD{
  M.~Alishahiha, M.~R.~M.~Mozaffar and A.~Mollabashi,
  ``Holographic Aspects of Two-charged Dilatonic Black Hole in AdS5,''
[arXiv:1208.2535 [hep-th]].
%%CITATION = arXiv:1208.2535%%
}

%\BhattacharyaZU
\lref\BhattacharyaZU{
  J.~Bhattacharya, S.~Cremonini and A.~Sinkovics,
  ``On the IR completion of geometries with hyperscaling violation,''
[arXiv:1208.1752 [hep-th]].
%%CITATION = arXiv:1208.1752%%
}

%\KunduJN
\lref\KunduJN{
  N.~Kundu, P.~Narayan, N.~Sircar and S.~P.~Trivedi,
  ``Entangled Dilaton Dyons,''
[arXiv:1208.2008 [hep-th]].
%%CITATION = arXiv:1208.2008%%
}

%\OgawaFW
\lref\OgawaFW{
  N.~Ogawa and T.~Takayanagi,
  ``Higher Derivative Corrections to Holographic Entanglement Entropy for AdS Solitons,''
JHEP {\bf 1110}, 147 (2011).
[arXiv:1107.4363 [hep-th]].
%%CITATION = arXiv:1107.4363%%
}

%\MyersED
\lref\MyersED{
  R.~C.~Myers and A.~Singh,
  ``Comments on Holographic Entanglement Entropy and RG Flows,''
JHEP {\bf 1204}, 122 (2012).
[arXiv:1202.2068 [hep-th]].
%%CITATION = arXiv:1202.2068%%
}

\Title{}
{\vbox{\centerline{On Holographic Entanglement Entropy of Charged Matter}
%\bigskip
%\centerline{}
}}
\bigskip

\centerline{\it  Manuela Kulaxizi${}^{1}$, Andrei Parnachev${}^{2}$ and Koenraad Schalm${}^{2}$}
\bigskip
\smallskip
\centerline{${}^{1}$ Universit\'{e} Libre de Bruxelles and International Solvay Institutes,} 
\centerline{ULB Campus Plaine CP231, B-1050 Bruxelles, Belgium }
\smallskip
\centerline{${}^{2}$Institute Lorentz for Theoretical Physics, Leiden University} 
\centerline{P.O. Box 9506, Leiden 2300RA, The Netherlands}
\smallskip

\vglue .3cm

\bigskip

\let\includefigures=\iftrue
\bigskip
\noindent
We study holographic entanglement entropy in the background of charged dilatonic black holes
which can be viewed as holographic duals of certain finite density
states of $\NN=4$ super Yang-Mills.
%KS
These charged black holes are distinguished in that they have
vanishing ground state entropy.
%DROP We consider the limit of vanishing temperature which corresponds to the vanishing ground state entropy.
%ENDKS
The entanglement entropy for a slab experiences a 
%KS 
second order 
%END KS
phase transition as the thickness of the slab is varied,
while the entanglement entropy for a sphere is a smooth function of the radius.
%KS
This suggests that the second scale introduced by the anisotropy of
the slab plays an important role in driving the phase transition.
%ENDKS
In both cases we do not observe any
%KS 
logarithmic 
violation of the area law indicative of hidden Fermi surfaces.
%ENDKS
We investigate how these results are affected by the inclusion of the Gauss-Bonnet term in the bulk gravitational action.
We also observe that such addition to the bulk action does not change the logarithmic violation of the
area law in the backgrounds with hyperscaling violation.
\bigskip

\Date{September 2012}

%\draftmode
\newsec{Introduction and summary}

\noindent Understanding strongly interacting compressible states of matter has been a subject of intense 
research (see e.g. \SachdevDQ\ for a recent review).
In this pursuit, it is natural to use the AdS/CFT correspondence, which provides an analytic
machinery of dealing with strongly interacting field theoretic systems.
A natural route involves starting with the well-understood duality between the $\NN=4$ Super Yang Mills
in $3+1$ dimensions and type IIB supergravity in $AdS_5\times S^5$ and deforming it by
turning on global charge density.
The simplest setup 
%KS
--- a truncation to Einstein-Maxwell theory --- 
%ENDKS
gives rise to the extremal RN-AdS black hole at vanishing temperature; fermion correlators in
this background exhibit poles which can be associated with (non)Fermi liquid behavior \refs{\LiuDM\CubrovicYE-\FaulknerWJ}.

% The near-horizon geometry of extremal RN-AdS black hole involves an
% $AdS_2$ factor and features
% finite entropy density, something that is at odds with the Fermi
% liquid physics.
%KS
The $AdS_2$ near-horizon geometry of the extremal RN-AdS black hole is
the bulk feature responsible for this exotic behavior. However, the
generic extremal black hole also has a finite entropy density at zero
temperature signalling that the groundstate is unstable.
%ENDKS
It is natural to ask whether there are finite density states,
%KS
preserving the $AdS_2$ factor,
%ENDKS
that do have vanishing entropy at zero temperature.
The answer is affirmative; turning on two out of three diagonal $U(1)$
global charges 
%KS 
in the IIB $AdS_5 \times S^5$ supergravity 
%ENDKS
gives rise to 
%KS
%the
a special class of charged
%ENDKS
 dilatonic black holes studied in \refs{\GubserQT-\GubserYB} (see also \DeWolfeUV\ for recent developments).
The ten-dimensional string metric is known and can be dimensionally reduced down to five dimensions.
The resulting 
%KS
near-horizon
%ENDKS
geometry is conformal to $AdS_2\times R^3$, but the conformal factor ensures that the entropy
at small temperatures vanishes linearly, $S\sim T$.

As shown in \GubserYB, fermion correlators in this background can be computed
analytically and exhibit the
%KS
%Fermi
sought-for 
(non-)Fermi
%ENDKS 
liquid behavior,
%KS
but now in an entropically stable system. In addition the system may carry
additional fermionic excitations ``behind the horizon''
\refs{\HuijseEF}.
%ENDKS 
A %useful 
%KS
potential
%ENDKS
probe of the presence of %KS
these ``hidden''
%END KS
Fermi surfaces is the entanglement entropy (EE).
As discussed in \refs{\WGK\SwingleBF\ZhangRW-\DSY}, in $d+1$ dimensional theory entanglement entropy of a ball of radius $R$ in the presence of the
Fermi surface behaves like $S_{ball,FS}\sim  R^{d-1} \log R$, which violates the naive area law, $S_{ball}\sim R^{d-1}$.  
In general, computing entanglement entropy in the interacting field theories is a very nontrivial exercise.
Fortunately, in the context of holographic duality, a simple prescription has been
formulated by Ryu and Takayanagi \refs{\RyuBV-\RyuEF}.
Recent work on holographic matter
with hyperscaling violation \HuijseEF\  (following \OgawaBZ; see also \refs{\ShaghoulianAA\DongSE\NarayanHK\KimNB\HashimotoTI\PerlmutterHE\CadoniUF\AmmonJE\BhattacharyaZU-\KunduJN}
for subsequent work on holographic entanglement in the presence of hyperscaling violation) identified a set of examples, labeled by the
dynamical critical exponent $z$, which exhibit such logarithmic violation.
All these examples involve the hyperscaling violation parameter set to be $\theta=d-1$ and feature
the low temperature behavior of the thermal entropy $S\sim
T^{(d-\theta)/z}\sim T^{1/z}$.
%KS
%I would move this to *****
% How general is this result? In this paper we try to address this question by
% repeating the calculation of \HuijseEF\ for more general class of theories.
% Namely, in addition to the Einstein-Hilbert gravitational action with a negative cosmological
% constant, we also consider a Gauss-Bonnet term in the bulk action.
% The prescription for computing entanglement entropy in the presence of such a term
% has been formulated in \refs{\FursaevIH\deBoerWK-\HungXB}, and we assume that it is not
% modified in the presence of additional matter in the bulk.
% The results are very similar to the Einstein-Hilbert case \HuijseEF; we describe the details 
% of our calculations in the Appendix.
%ENDKS

It has been pointed out in \HartnollWM\ that in order for models with generic values of 
$z$ and $\theta$ to holographically describe  Fermi
liquids, one should take the limit $z\ra\infty,~\theta\ra -\infty$ with the limit $\theta/z$ held fixed.
This is necessary in order to have a finite spectral density at finite values of momenta.
Interestingly, the %KS
entropically stable
charged dilatonic black holes
%metric
%ENDKS
studied in \GubserYB\ precisely fall in this class.
As we have already remarked, it also exhibits specific heat which vanishes linearly in 
temperature.
As pointed out in \HartnollWM, the simplest computation of entanglement of a belt in this geometry
involves a phase transition and the resulting hypersurface in the bulk becomes disjoint when the
belt becomes sufficiently thick.
%KS 
The latter phase 
%ENDKS
of course signals the simple area law for the entanglement entropy.

In this paper we compute the entanglement entropy of a ball, and observe that
contrary to the slab case, there is no phase transition to the disjoint configuration.
However, the numerical results indicate that in the limit of a large size, the
entanglement entropy of a ball still exhibits the area law. 
%KS
The explicit absence of a phase
transition for ball-entanglement-entropy is odd compared to the belt
configuration. 
%One would expect the transition to be universal in
%terms of the ratio of the size of the entanglement surface and the
%charge density. 
%The anisotropic belt configuration has a second scale
%however, and this allows one to understand the phase transtion. Rather
%than an ovoid distortion of the sphere, 
One may think of the belt
as the limit of two concentric balls with large radii but fixed
difference. Geometrically, the belt phase transition is then of the
Gross-Ooguri type for concentric Wilson loops
\refs{\GrossGK\ZaremboBU\OlesenJI\KimTD\HirataJX-\PakmanUI}.
When the difference between the radii is small, the minimal embedding
surface is a half-torus; when the difference is large, the minimal
embedding is two concentric spheres.
%ENDKS

We also investigate entanglement entropy in the presence of higher derivative
(Gauss-Bonnet) term in the gravity action
%.
%KS *****
%In this paper we try to address this question by
to address how general these results are.
% repeating the calculation of \HuijseEF\ for more general class of theories.
% Namely, in addition to the Einstein-Hilbert gravitational action with a negative cosmological
% constant, we also consider a Gauss-Bonnet term in the bulk action.
The prescription for computing entanglement entropy in the presence of such a term
has been formulated in \refs{\FursaevIH\deBoerWK-\HungXB}, and we assume that it is not
modified in the presence of additional matter in the bulk.
% The results are very similar to the Einstein-Hilbert case \HuijseEF; we describe the details 
% of our calculations in the Appendix.
In the holographic Gauss-Bonnet gravity we restrict our attention to the EE of a slab in 
the geometry with hyperscaling violation (which described the near horizon region
the the Einstein-Hilbert case).
Already here we encounter a surprise: the connected entangling surface does not end on
the boundary for large values of the thickness $\ell$ of the slab; it can instead approximate 
the boundary but never get there.
We expect that completing the geometry would lead to the same picture as that for
the ball entanglement entropy in the Einstein-Hilbert case: no phase transition and
a connected surface in the bulk defined for all large values of $\ell$.

The %KS
%rest of the
%ENDKS
paper is organized as follows.
In the next Section we review both the 5-dimensional dilatonic black hole
and its 10-dimensional lift and show that entanglement entropy calculation is
equivalent in both settings.
In Section 3 we compute entanglement entropy of simple geometries
and observe that the calculations in the case of a belt and and in the case of a ball
are very different.
In the former case there is a maximal value of the belt thickness, which can
support a curved entangling surface; beyond that value there is a phase
transition to a disjoined entangling surface.
In the case of a ball, the entangling surface in the bulk stays connected and arbitrary
large values of the ball radius $R$ are supported.
However, the entanglement entropy exhibits an area law in both cases.
In Section 4 we study EE in the case of Gauss-Bonnet gravity and comment on the
differences with the Einstein-Hilbert case.
We discuss our results in Section 5. 
%KS
In the appendix we present
%ENDKS 
some details of EE in Gauss-Bonnet gravity computed in the holographic
duals of states with hyperscaling violation.

{\bf Note Added:} When this paper was completed \AlishahihaAD\ appeared where the entanglement entropy of 
a straight belt in the five-dimensional black hole geometry was computed. This partially overlaps with Section 3.

\newsec{Entanglement entropy and dimensional reduction}

\noindent As pointed out in the introduction, one of the characteristic properties of systems with a Fermi surface,
is the logarithmic violation of the area law of entanglement entropy \refs{\WGK\SwingleBF\ZhangRW-\DSY}. 
Such logarithmic violations were recently found \refs{\HuijseEF\OgawaBZ-\ShaghoulianAA} in a class of holographic,
albeit singular (with exception the case $\theta>0$ and $z=1+{\theta\over d}$ in $d+2$ bulk spacetime dimensions\foot{We thank Edgar Shaghoulian for correcting a mistake in this statement in the earlier version of the article.}), geometries with flux.
It was subsequently realized that these class of geometries additionally exhibit other characteristic properties
of Fermi surfaces, such as the scaling of the entropy with temperature at low temperatures, \ie, $S\sim T^{1\over z}$ 
\HuijseEF. 

The geometries in question are characterized by a critical exponent $z$ and a hyperscaling violation 
exponent $\theta$ and are described by 
\eqn\hyperviod{ds^2={1\over r^2}  \left(- { dt^2\over r^{{2d(z-1)/(d-\theta)}} } +g_0 r^{{2\theta\over (d-\theta)}} dr^2+\sum_{i=1}^d dx^2\right) \, ,}
where the UV boundary is located at $r=0$ and the IR corresponds to
large values of $r$. 
They 
can be found as solutions of the
%KS 
%DROP Einstein 
%ENDKS
field equations emanating from 
\eqn\Lhyper{\LL=R-Z(\Phi) F^2-(\p\Phi)^2-V(\Phi)\,}
where the dilatonic scalar field $\Phi$ and the Maxwell field $A$ are additionally turned on.
The parameter $g_0$ which appears in \hyperviod\ is related to the IR behavior of the potential $V(\Phi)$. 

This class of models has been extensively discussed in 
\refs{\TaylorTG\GoldsteinAW\CharmousisZZ\CadoniKV-\GouterauxCE}.
The logarithmic violation of the area law in entanglement entropy is realized for $\theta=d-1$.
Scale transformations act on these geometries \HuijseEF\
\eqn\scaleact{\eqalign{\overrightarrow{x}&\rightarrow \Lambda  \overrightarrow{x}\cr
t&\rightarrow \Lambda^z t\cr
ds&\rightarrow \Lambda^{\theta\over d}ds\cr
r&\rightarrow \Lambda^{{(d-\theta)\over d}} r
}}
and lead to the following behavior for the thermal entropy $S\sim T^{{(d-\theta)\over z}}$,
explicitly verified in \OgawaBZ\ and \HuijseEF.
%KS
One recognizes how
the hyperscaling violation exponent $theta$ parametrizes the effective
dimensional reduction.
%ENDKS

An interesting special case of \hyperviod\ involves the double scaling limit 
\eqn\doublescalinglimit{z\rightarrow \infty \qquad \qquad -{\theta\over z}\equiv \eta>0\, ,}
with $\eta$ fixed. As emphasized in \HartnollWM\ several physical quantities behave as expected
for Fermi liquids in this limit. The spacetime metric becomes conformal to $AdS_2\times R^{d}$
\eqn\adsconfhyper{ds^2={1\over r^2} \left(-{dt^2\over r^{2d\over \eta}}+g_0 {dr^2\over r^2}+\sum_{i=1}^d dx_i^2\right)\, ,}  
while the thermal entropy vanishes like $S\sim T^{\eta}$ for small temperatures. 
More importantly the spectral density remains finite at low energies and finite momenta \HartnollWM.

%In general, solutions of the form \hypervio\ are better thought of as describing the IR region
%of asymptotically $AdS_{d+2}$ spacetime at finite chemical potential.

Here we are interested in a specific example of \adsconfhyper, the
dilatonic black hole in $AdS_5$
recently explored in \refs{\GubserQT-\GubserYB}. This black hole is usually referred to as the two-charge black hole,
because %KS
it arises as a truncation of IIB supergravity on $AdS_5\times S^5$
where two of the three $U(1)$ charges 
% of maximally gauged supergravity
%ENDKS
are equal and non-vanishing while the 
third one is zero \CveticXP. 

%KS
After this truncation one obtains 
%DROP Consider
%ENDKS
the Lagrangian
\eqn\LLgubser{\LL=R-{1\over 4}e^{4\alpha}F_{\mu\nu}^2-12 \left(\p_\mu\alpha\right)^2+{1\over L^2}\left(8e^{2\alpha}+
4 e^{-4\alpha}\right))}
with $\alpha$ a scalar field and $F_{\mu\nu}$ the field strength of the Maxwell field $A_{\mu}$.
The extremal black hole metric
\eqn\bhmetric{  ds^2 = e^{2 A} \left( - h dt^2 +dx^2\right)+{e^{2 B}\over h} dr^2  }
with
\eqn\defab{\eqalign{ A&=\ln{r\over L} +{1\over3} \ln\left(1+{Q^2\over r^2}\right), \cr
B&=-\ln{r\over L} -{2\over3} \ln\left(1+{Q^2\over r^2}\right) \cr
h&= {(r^2+2 Q^2) r^2\over (r^2+Q^2)^2}  
}}
together with 
\eqn\defrest{A_\mu dx^\mu={\sqrt{2}Q r^2\over L (r^2+Q^2) } dt,\qquad\qquad \alpha={1\over 6}\ln{\left(1+{Q^2\over r^2}\right)}}
form a solution of the equations of motion coming from \LLgubser. In these coordinates, large $r$ corresponds to the near-boundary 
UV region while the black hole horizon is located at $r=0$. To be
precise, $r=0$ is actually a naked singularity.
%KS 
Any finite temperature will cloak the singularity, however, so it is
of the ``good'' type.
%ENDKS
In the near-horizon region, \bhmetric\ is conformal to $AdS_2\times R^3$ \GubserYB
\eqn\confadsa{ds^2=\left({r\over Q}\right)^{{2\over 3}} \left(-{2 r^2\over L^2}dt^2+{L^2\over 2 r^2}dr^2+{Q^2\over L^2}\sum_{i=1}^3 dx_i^2\right)\, .}
With a change of variable from $r$ to $\tilde{r}$ such that ${r\over Q}={1\over \sqrt{2} \tilde{r}^3}$ the IR metric reduces to
\eqn\confadsr{  ds^2 ={1\over \tilde{r}^2} {Q^2\over 2^{1\over 3} L^2} \left( -{dt^2\over \tilde{r}^6}+{9 L^4\over 2 Q^2}{d\tilde{r}^2\over \tilde{r}^2} +\sum_{i=1}^3 dx_i^2 \right) \,.}
Thus \confadsr\ is of the form \adsconfhyper\ with $g_0={9 L^4\over 2 Q^2}$ and $\eta=1$.

%KS 
%DROP It is an interesting fact that \LLgubser\ can be obtained from a consistent tuncation of type IIB supergravity
% DROP with $Q_1=Q_2=Q$ and $Q_3=0$ \CveticXP.
By the definition of consistent truncation, this solution can be
lifted to a solution of the full IIB supergravity with $Q_1=Q_2=Q$ and $Q_3=0$ \CveticXP.
This 
%ENDKS
ten dimensional solution is 
\eqn\metricuplift{\eqalign{ds^2&=\sqrt{\Delta}H^{-{2\over 3}} \left[e^{2 A}\left(-hdt^2+\sum_{i=1}^3dx_i^2\right)+
{e^{2 B}\over h}dr^2\right]\cr
&+{L^2 H\over\sqrt{\Delta} }\sum_{i=1}^2  \left[d\mu_i^2+\mu_i^2 \left(d\phi_i+{Q\over L^2 H}
(1-H)dt\right)^2\right]+{L^2\over \sqrt{\Delta}}  \left(\cos^2{\theta_1}d\theta_1^2+\sin^2{\theta_1}d\phi_3^2\right)\cr
F_5&=G_5+\star G_5\qquad G_5=dB_4\qquad B_4=-{r^4\over L^4}\Delta dt\wedge d^3 x-{Q^3\over L^2}\sum_{i=1}^2 \mu_i^2d\phi_i\wedge d^3x 
\,.}}
where
\eqn\defs{H=1+{Q^2\over r^2},\quad \Delta=H(\cos^2\theta_1+H \sin^2\theta_1)\quad }
and 
$h,\,A,\,B$ are given in \defab. Here $\mu_1$ and $\mu_2$ are defined as 
\eqn\anglesdef{\mu_1\equiv\cos{\theta_1}\cos{\theta_2},\qquad \mu_2\equiv\cos{\theta_1}\sin{\theta_2}\,.}
where $\theta_1,\theta_2$ are coordinates on the $S^2$ inside the $S^5$which is parametrized by 
$\left(\theta_1,\,\theta_2,\,\phi_1,\,\phi_2,\,\phi_3\right)$.
The solution describes $N$ coincident $D3$ branes, rotating with equal angular momentum in the two 
out of the three possible planes of rotation transverse to the $D3$-brane world-volume. 
%The singularity...

As explained in the introduction, our main focus will be on entanglement entropy. 
A natural question to ask is whether the computation of entanglement entropy in the ten dimensional 
geometry \metricuplift\ is equivalent with the one in five dimensions. In the simple case of p-branes,
\KlebanovWS\ and \DongSE\ showed that the two computations, in the ten dimensional setup and
in the dimensionally reduced geometry, agreed as long as the former was done in string frame 
and the latter in the Einstein frame. 
%This was clearly a strong confirmation of the proposal of [RyuTakayanagi].
%However since a definition for entanglement entropy as a string theory object is lacking, there is no 
%a priori guarantee that the two computations will always agree. 
In what follows, we will show that this is also true for \metricuplift\ and \bhmetric.
 
Let us consider the entanglement entropy of some arbitrary connected region 
and parametrize the entangling surface by a single function $r(x_1,\, x_2,\, x_3)$. 
The induced metric for a surface in the geometry of \metricuplift\ which asymptotes 
to the entangling surface under consideration is 
\eqn\eeinduced{\eqalign{ds^2_{EE}&=\sqrt{\Delta} H^{-{2\over 3}} e^{2 A}\left[\delta_{ij}+
{e^{2B-2 A}\over h}(\p_i r)(\p_j r)\right]dx^i dx^j+\cr
&+{L^2 \over\sqrt{\Delta}}\left[Hc_1^2d\theta_2^2+H c_1^2\left(c_2^2d\phi_1^2+s_2^2d\phi_2^2\right)+s_1^2 d\phi_3^2\right]+
{L^2\sqrt{\Delta}\over H} d\theta_1^2  
}}
where we introduced the notation $c_i=\cos{\theta_i},\, s_i=\sin{\theta_i}$
and used the definition of $\Delta$ in \defs\ to simplify the expression.
The entropy functional is proportional to the volume of the induced metric, which integrated
over the five dimensional compact space yields
\eqn\eelang{\sqrt{g_{EE}}= Vol_5\,  e^{3 A} \sqrt{ \det{\left[\delta_{ij}+{e^{2B-2A}\over h} (\p_i r)(\p_j r) \right]}} \,.}
where $Vol_5=L^5\Omega_5$ is the volume of the five dimensional sphere of radius $L$.
Finally,  the entanglement entropy computed with the ten-dimensional metric \metricuplift\ is given by 
\eqn\eeten{S^{EE}_{slab} = {1\over G_N^{5}} \int d^3x e^{3 A} \sqrt{ \det{\left[\delta_{ij}+\p_i r\p_j r {e^{2B-2 A}\over h} \right]}} }
where we used the relation ${1\over G_N^{5}}={Vol_5\over G_N^{10}}$.
As expected, \eeten\ precisely coincides with the entropy functional computed in the dimensionally
reduced metric \bhmetric.
%KS
Note that in the ten-dimensional metric the singularity is
that of a conventional rotating black brane. The equivalence thus
guarantees that the apparent naked
singularity in the five-dimensional extremal black hole metric
\bhmetric~is innocuous and does not affect the results. 
%ENDKS

In summary, computing entanglement entropy in the ten-dimensional geometry is completely
equivalent to computing it in the five dimensional geometry.

\newsec{Entanglement entropy for the dilatonic black hole}

\noindent In this Section we will compute holographic entanglement entropy in the theory described
by the metric \bhmetric.
In the following we set $L=1$.
It will be useful to write the near-horizon limit of the metric \bhmetric\ in the following form
\eqn\confads{  ds^2 =  2^{-{5\over3}} Q^{-{2\over 3}} z^{-{2\over 3} } \left( {-dt^2+dz^2\over z^2} +  \sum_{\mu=1}^3 d\tilde x_\mu^2 \right) }
where $\tilde x_\mu=\sqrt{2} Q x_\mu$ and $z=1/2r$.
Note that the metric \confads\ belongs to the class of metrics studied in  \HuijseEF\
with a particular value of dynamical exponent $z=\infty$.
We start by placing a system in the large 3-dimensional box of size $L_x$, which serves as an IR regulator.
We can now compute  the entanglement entropy of a slab, specified by $0\leq x_1\leq \ell$, $0\leq x_2,x_3\leq L_x$,
where $\ell \ll L_x$.
The entangling surface can be parameterized by a single function $x_1(z)$ with the boundary conditions
$x_1(0)=0,\ell$.
The entropy functional takes the form
\eqn\eeslab{   S_{slab} =  L_x^2 \int {dz\over z} \sqrt{ {1\over z^2} + (x')^2 }   } 
where $x\equiv \sqrt{2} Q x_1$, the prime denotes derivative with respect to $z$ and the integral needs to be appropriately
regularized.
In \eeslab\ and in the rest of the paper we omit an overall numerical factor, together with
a factor of $L^2/\ell_p^2$, which roughly measures the number of degrees of freedom (in 
the case of the holographic dual of a conformal field theory such a
factor is proportional to the central charge).
%KS
The
%ENDKS
equation of motion that results from \eeslab\ can be written as
\eqn\eomslab{  {x'\over z\sqrt{z^{-2}+ (x')^2} } = {1\over z_0}   }
where $1/z_0$ is a constant of motion; the value of $z=z_0$ corresponds to the
turning point of the curve in the bulk where $x'=\infty$. 
By varying the value of $z_0$ one can change the shape of the curve.
The value of $\ell$ as a function of $z_0$ can be computed via
\eqn\ellznot{ \sqrt{2} Q \ell = 2 \int dz x' = 2 \int_0^{z_0}  {dz\over z}  {1\over \sqrt{(z_0/z)^2-1 }  } = \pi.}
The key feature of this result is its independence of the value of $z_0$: the curved hyper\-surface in 
the bulk which approximates to the boundaries of the slab on the $AdS$
boundary can only do so for a fixed
thickness of the slab $\ell$. 
We are, of course, interested in the behavior of the EE as $\ell$ is taken to be large.
At this point it is worth recalling that there is always a trivial solution $x'=0$: this becomes the
only available solution for $Q \ell\neq\pi/\sqrt{2}$.
It is clear that the entanglement entropy for such a solution is independent of $\ell$.
(this feature of the geometry \bhmetric\ has been pointed out before; see e.g.  \HartnollWM.)

 When the full geometry \bhmetric\ is considered, we might expect the picture to be the following.
 For sufficiently large $\ell$ the hyper\-surface in the bulk should probe the infrared limit of the geometry \bhmetric,
 described by \confads.
 Therefore the discussion of the previous paragraph applies: there is a maximal value of $\ell=\ell_{crit}$ which 
 corresponds to the curved solution; beyond it only the trivial $x'=0$ solution exists.
 The transition between the two configurations is second order, since the curved solution asymptotically
 approaches the trivial one as  $\ell\ra \ell_{crit}$.
 For sufficiently small $\ell$ the entangling surface probes the UV (small $z$ region) of the geometry, and
 we expect $\ell$ to be a smooth function of $z_0$ with a limit $\ell(z_0=0)=0$.
 To check these expectations we repeat the calculation above for the full geometry \bhmetric.
 The EE functional is now given by
 \eqn\eeslabb{   S_{slab} =  L_x^2 \int dz \, e^{2A} \sqrt{ e^{2B}/h + e^{2 A} (x')^2 }   } 
The analog of \ellznot\ is now given by
\eqn\ellznotb{  Q \ell = 2 \int_{\tau_0}^\infty {d \tau \over \sqrt{ \tau^2 (2+\tau^2) \left( \tau^2 (1+\tau^2)^2 \tau_0^{-2}  (1+\tau_0^2)^{-2} -1\right) } } } 
where $\tau=r/Q, \tau_0=r_0/Q$.
The value of this integral as a function of $\tau_0$ is plotted in Fig. 1.
\ifig\loc{$\ell Q$ as a function of $\tau_0$ computed using eq. \ellznotb.}
{\epsfxsize3.3in\epsfbox{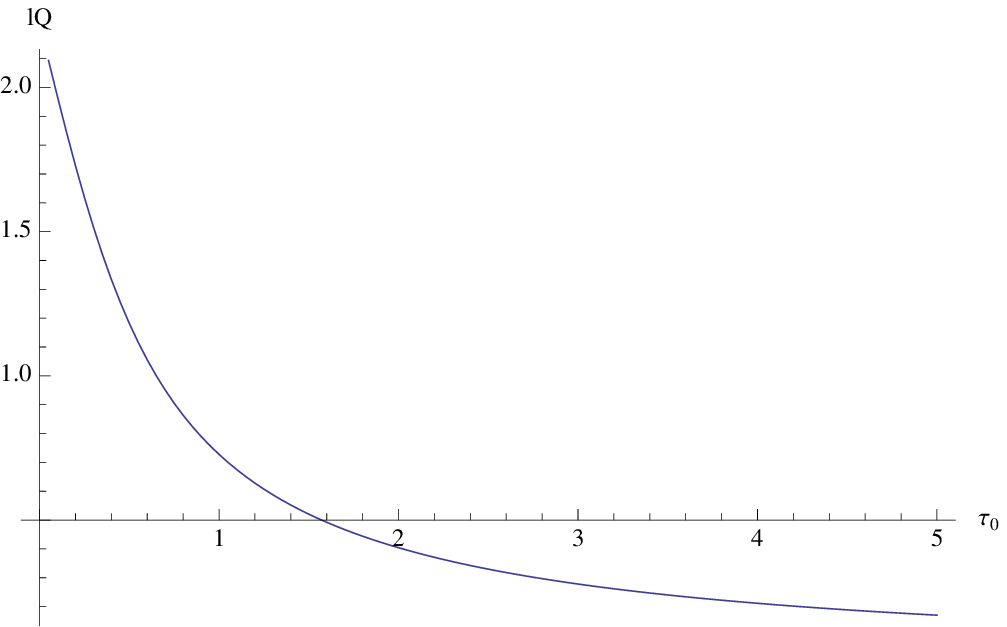}}
\ifig\loc{$\Delta S$ as a function of $\ell Q$.}
{\epsfxsize3.3in\epsfbox{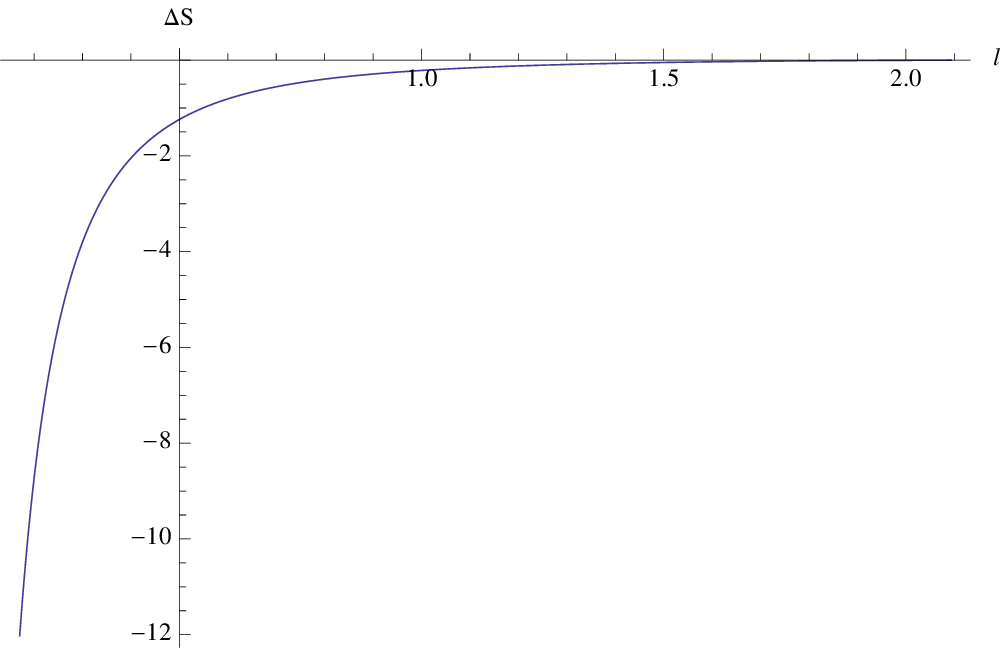}}
The figure is in accord with the general discussion above.
In the UV region, $\tau_0\gg1$, the length of the interval monotonically decreases to zero.
On the other hand, there is a limiting value of $\ell$ as the position of the tip approaches the
horizon at $\tau=0$.
Note that this limiting value is  equal to $\ell_{crit}=\pi/\sqrt{2} Q$, in accord with the near-horizon 
analysis.
We check that for $\ell\le  \ell_{crit}$ the curved solution is preferred (gives the leading contribution to the EE),
and the difference vanishes in the limit $\ell\ra \ell_{crit}$.
(see Fig. 2).

%KS
\bigskip
%ENDKS

The next step is computing the entanglement entropy for a sphere, defined by 
\eqn\defsphere{  \sum_{\mu=1}^3 x_\mu^2 ={ \rho^2\over 2 Q^2} \leq R^2   }
We start by considering the %KS
near-horizon
%ENDKS
metric \confads.
The entangling surface in the bulk can be parameterized by a single function $z(\rho)$; the EE functional is given by
\eqn\eesphere{  S_{sphere} =  \int  {d\rho\over z} \rho^2 \sqrt{ (z')^2/z^2+1}.    }
%KS
The equation
%ENDKS
 of motion is now second order:
\eqn\eomsphere{  {\p\over \p \rho} \left( {\rho^2 z'\over z^3\sqrt{(z')^2/z^2+1}}\right) =   -{\rho^2 (z^2+2 (z')^2)\over z^4\sqrt{(z')^2/z^2+1}} }
We can integrate this numerically; the result is given by the red curve  in Fig. 3. 
%\ifig\loc{$z(\rho)$  computed using eq. \eomsphere.}
%{\epsfxsize3.3in\epsfbox{zrho.eps}}
%
The surprising feature of eq. \eomsphere\ is the absence of the
solutions approaching the boundary at $z=0$ at finite $\rho$.
Such a behavior has not been observed in the examples studied in \HuijseEF.
This behavior is evident in Fig. 3. (the red curve never intersects the $z=0$ line).
Also, unlike the slab case, there are no trivial solutions
$\rho(z)=const$.

%KS
The near-horizon approximation is clearly not reliable and we must
% As a next step,
we analyze the EE of a sphere % for
in
the full diatonic black hole metric \bhmetric.
%ENDKS (moved figures to end of paragraph)
The EE functional now takes the form
\eqn\eesphereb{  S_{sphere} = \int  {d\rho\over z} \rho^2  e^{2 A(z)} \sqrt{e^{2 A(z)}+ {e^{2B(z)} (z')^2\over 4 z^4 h(z)}}    }
where $A(z), B(z),h(z)$ are determined by \defab\ with the
substitution $z=1/2r$.
The 
equation of motion for $z(\rho)$ that follows from \eesphereb\ is second order and has to be dealt with
numerically.
A representative solution is shown in Fig 3 (blue curve).
The red curve corresponds to the solution in the near horizon geometry.
%KS DROPThe crucial difference between the two curves is that the entangling surface in the near-horizon geometry
%never reaches the boundary at $z=0$, while 
%the 
%ENDKS
The modification of the EOM in the full geometry 
ensures that the hypersurface %KS 
now 
%ENDKS
reaches the boundary at finite $\rho$.
\ifig\loc{$z(\rho)$  for the near horizon (red) and full BH metric (blue)}
{\epsfxsize3.3in\epsfbox{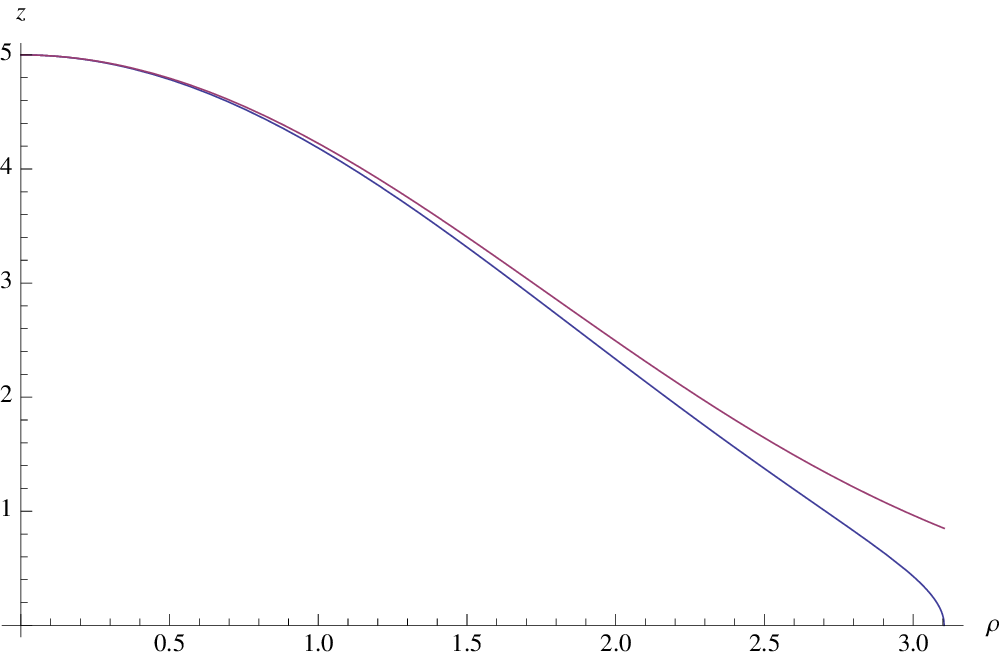}}
\ifig\loc{$R$  as a function of $z_0$ for the full BH metric}
{\epsfxsize3.3in\epsfbox{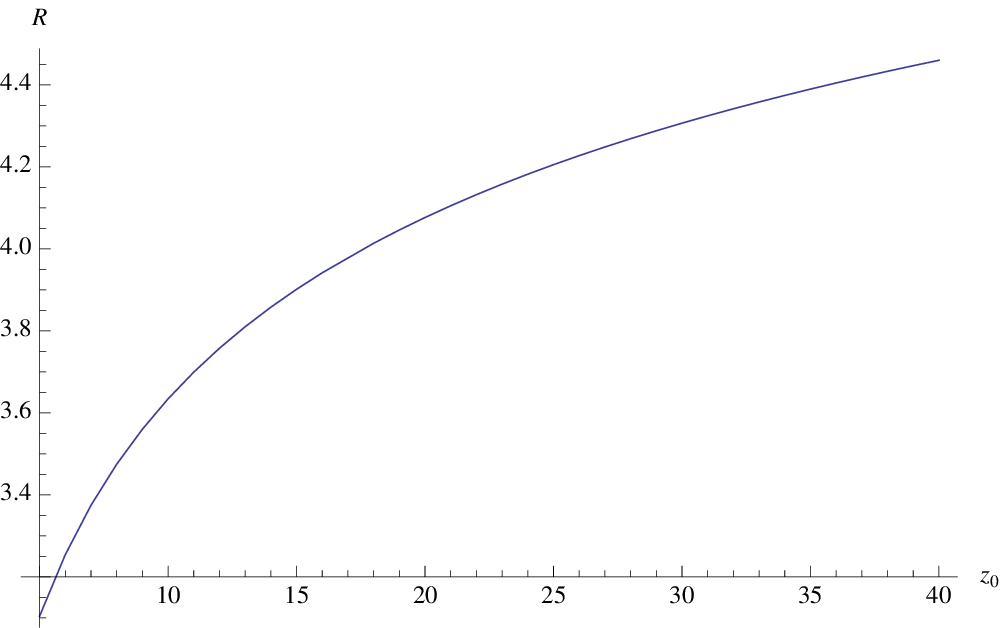}}

We can %KS
now
%END KS
compute the radius of the entangling surface at the boundary as a function of $z_0$
numerically.
The result is shown on Fig 4.
Finally we are in position to compute the value of entanglement entropy.
(Similar computations have recently appeared in \LiuEE\ and \MyersED; we will follow the procedure outlined in \PakmanUI,
which the reader is encouraged to consult for technical details.)
The entanglement entropy of a ball is UV divergent and a suitable regularization is necessary.
The UV divergent terms are governed by the asymptotically $AdS_5$ metric; these terms in the
conformal case have been computed in \RyuEF:
\eqn\divergenball{ S_{ball,divergent}= {R^2\over 8 \epsilon^2} +{1\over 2} \log\epsilon  }
here $\epsilon\ll1$ is the cutoff on the radial coordinate $z$.
An extra factor of $1/2$ in the first term in \divergenball\ compared to eq. (7.10) in \RyuEF\ appears
due to the factor of two in the definition of $z$ in our paper.
\ifig\loc{$S_{UV finite}/R^2$ as a function of $R Q$; red curve corresponds to $S_{UV finite}/Q^2R^2=0.25-0.78/Q^2R^2$ fit.}
{\epsfxsize3.3in\epsfbox{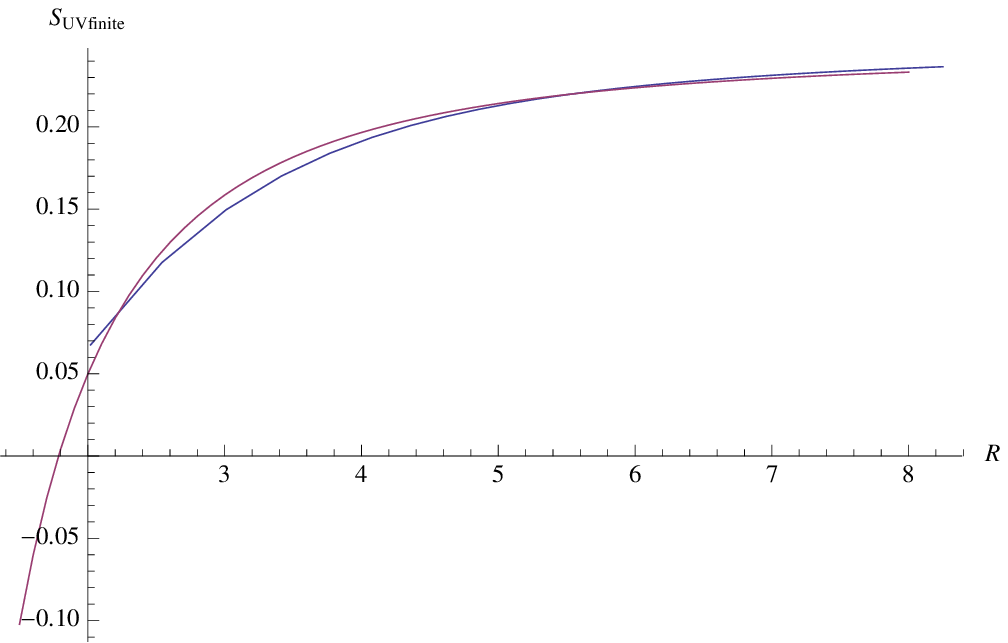}}
All we need is to evaluate  \eesphereb\ on the solution of equation of motion, and subtract the UV divergent
term \divergenball.
We are interested in the deviation of the resulting finite quantity, $S_{UV finite}=S_{ball}- S_{ball,divergent}$
from the area law.
In Fig. 5 we plot the normalized UV finite part of the entanglement entropy.
% The result implies that
%KS
One observes directly that at large $R$ the behavior of entanglement entropy of a ball is governed 
by the area law. The equation of motion does not appear to have a
second distinct class of solutions and the result implies that the
ball-entanglement entropy does not experience a phase transition at
smaller radius to a phase differing from the naive area law.
%ENDKS 

\newsec{Gauss-Bonnet and the EE of a belt}

\noindent Here we compute the entanglement entropy of a belt in Gauss-Bonnet gravity.
We use the near horizon geometry of eqs. \confads\ which is a solution to the equations of motion
following from dilatonic Gauss Bonnet gravity with flux (see the appendix for more details).
Entanglement entropy in Gauss-Bonnet gravity is given by extremizing the action functional \refs{\deBoerWK-\HungXB}
\eqn\acsig{  S= \int d^3x \sqrt{\det G_\Sigma } \left(1+ \lambda R_\Sigma \right) }
where $\lambda$ is the Gauss-Bonnet coupling constant whereas $G_\Sigma$ and $R_{\Sigma}$
refer to the induced metric and the induced scalar curvature \foot{To make the variational problem well--defined
a boundary term should be added to \acsig. This term only affects the leading 
(proportional to $\lambda$) divergent term in the entanglement entropy, e.g. see \HungXB, 
it will therefore not affect the results of this section. }. 

Consider the slab, specified by $0\leq x_1\leq \ell$ and $x_2,\,x_3\leq L_x$ with $L_x>>\ell$ as in section 3.
The entangling surface is parametrized by a single function $x_1(z)$ and gives rise to the induced
geometry
\eqn\eebeltgb{ds^2=2^{-{5\over 3}}Q^{-{2\over 3}}z^{{2\over 3}} \left[\left({1\over z^2}+\dot{\tilde{x}}_1^2(z)\right) dz^2+\sum_{i=2}^3d\tilde{x}_i^2\right]\,,}
where the dot implies differentiation with respect to the variable $z$ and $\tilde{x}_i=\sqrt{2}Q x_i$.
The action functional \acsig\ reduces in this case to
\eqn\slabactiongb{ S_{slab}=2^{-5}Q^{-2} L_x^2 \int {dz\over z} \sqrt{{1\over z^2}+\dot{x}^2} \left(1-{2\lambda z^{{2\over 3}}
\left(1+7 z^2 \dot{x}^2+6 z^3 \dot{x}\ddot{x}\right)  \over 9(1+z^2 \dot{x}^2)^2} \right)\, ,} 
where we set $x\equiv \sqrt{2}Q\tilde{x}_1$.
The equation of motion folowing from \slabactiongb\ takes the form
\eqn\eqmgb{{\dot{x}\over \sqrt{1+z^2 \dot{x}^2}}-{2\over 9} \lambda z^{{2\over 3}} {\dot{x} \over \left(1+z^2 \dot{x}^2\right)^{{3\over 2}}}={1\over z_0}}
where $z_0$ is a constant of motion related to the width of the slab $\ell\equiv {\tilde{\ell}\over \sqrt{2}Q}$ through
\eqn\ldef{{\tilde{\ell}\over 2}= \int_{z_0}^0 dz \dot{x}} 
It is convenient to define a dimensionless variable $y={z\over z_0}$ which takes values in $y\in \left[0,1\right]$
and express \eqmgb\ as follows
\eqn\eqmgby{ {x'\over \sqrt{1+y^2 x'^2}} -
{2\over 9} \lambda z_0^{{2\over 3}} {x'\over \left(1+y^2 x'^2 \right)^{{3\over 2}}}-1=0\, ,}
where now primes indicate differentiation with respect to the new variable $y$. 
%Note that \eqmgby\ depends on $z_0$ only through the product $\lambda z_0^{{2\over 3}}$
%which we refer to as $B\equiv \lambda z_0^{{2\over 3}}$ in the following. 
Eq. \eqmgby\ shows that the presence of the Gauss-Bonnet coupling strongly affects the physics of the entanglement entropy.
In particular, $x'(y)$ now depends on $z_0$ through $B\equiv \lambda z_0^{{2\over 3}}$. As a result the length 
of the slab $\tilde{\ell}$, which in $y$ coordinate is given by
\eqn\lb{{\tilde{\ell}\over 2}=\int_0^1 dy x'(y) \, ,}
is not  anymore independent of $z_0$. Note that we have taken $z_0\geq 0$  to make contact with the $\lambda=0$
limit.

To proceed we need to solve \eqmgby\ for $x'(y)$. Obvisouly, the disconnected solution, $x'=0$
is always a solution. There also exist three connected solutions to \eqmgby\ but, as commonly observed in
Gauss-Bonnet gravity, only one of them asymptotes to the connected solution when $\lambda$ 
vanishes\foot{It would be interesting to study all possible solutions in detail in the spirit e.g. of \OgawaFW.}.
We choose this solution for $x'$ and proceed to examine its behavior close to the boundary $y=0$.
We find that for all $B\geq {9\over 2}$ $x'(y)$ diverges as $y$ approaches the boundary
\eqn\solba{x'=\left\{ \eqalign{
&{\sqrt{-9+2 B} \over 2 y} +{\sqrt{2} B^{3/2}\over -27+6 B}+\OO(y)\qquad B>{9\over 2}\cr
&{1\over y^{2\over 3}}+{1\over 2}+{3 y^{2\over 3}\over 8}+\OO(y^{4\over 3})\qquad\qquad\quad\,\,\, B={9\over 2}}
\right. }
whereas it goes to a constant for $B< {9\over 2}$
\eqn\solbb{x'={9\over 9-2 B} +{9\over 2} 3^{1\over 3} {-3+2 B\over \left(-9+2 B\right)^4} y^2+\OO(y^4) \qquad B<{9\over 2}}
Hence, a solution which asymptotes to the entangling surface only exists for $B<{9\over 2}$. A natural question 
to ask is whether this result changes when one considers the full solution and not just the near horizon geometry.
We leave this to future investigation.

It is instructive to also consider the behavior of the solution close to the turning point $y=1$. 
One essentially finds that the solution becomes complex for $B<-{9\over 4}$, in other words it ceases to exist.
For $B>-{9/4}$ $x'$ diverges at $y\sim 1$ and the solution tends to a constant value.
We will therefore restrict to values of $B$ lying in the rgion $-{9\over 4}\leq B < {9\over 2}$.

Substituting this solution into \lb\ and performing the integration numerically shows that for $B\in \left(-{9\over 4} ,\, {9\over 2}\right)$, the length of the slab $\tilde{\ell}$ varies from $\tilde{\ell}\in \left(0.4,\, {3\pi}\right)$. However, since $z_0\geq 0$, 
the sign of $B$ is completely determined by the sign of the Gauss-Bonnet coupling $\lambda$ which according to the null energy condition (see appendix) is bounded above by $\lambda\leq {9\over 2}$. This implies that both signs are allowed for $B$. It would be interesting to see if other constraints, such as those discussed in \refs{\BriganteNU\BriganteGZ\BuchelTT-\HofmanUG}, can uniquely fix the sign of $\lambda$.

The length $\tilde{\ell}$ as a function of $B$ for positive values of $B$ is plotted in Fig. 6. 
\ifig\loc{$\tilde{\ell}$ as a function of $B\equiv \lambda z_0^{2\over 3}$ for $B\geq 0$. A solution only exists for $\tilde{\ell}<3\pi$.}
{\epsfxsize3.3in\epsfbox{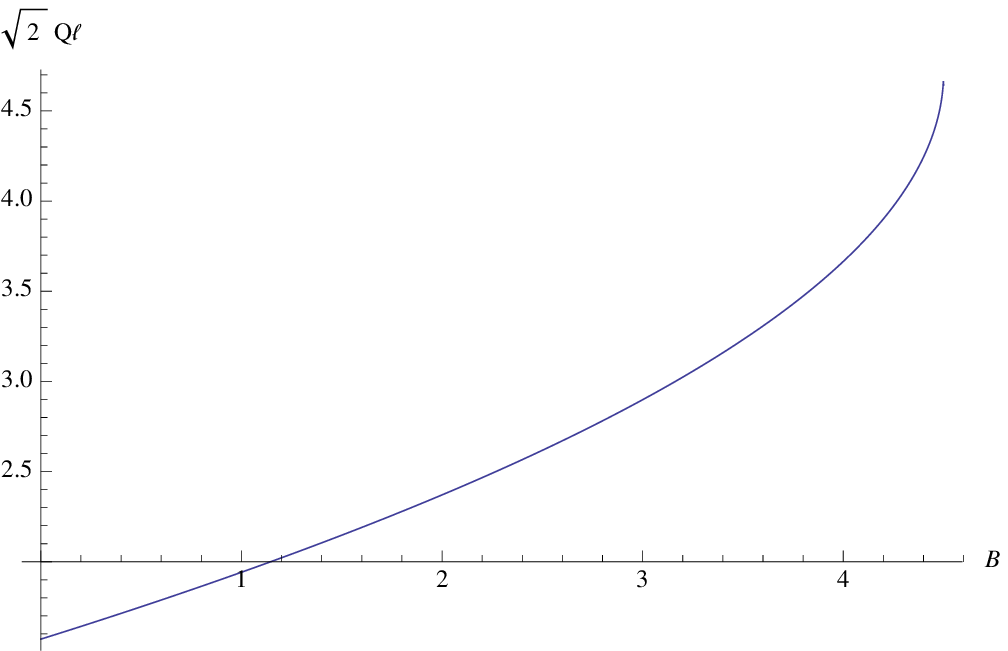}}
\noindent Fig. 7 instead shows $\tilde{\ell}$ for negative values of $B$. The two curves join at the point $B=0$ where 
the result of section 3 is recovered; the connected solution exists only for $\tilde{\ell}={\pi}$. 
\ifig\loc{$\tilde{\ell}$ as a function of $B\equiv \lambda z_0^{2\over 3}$ for $\lambda\leq 0$.}
{\epsfxsize3.3in\epsfbox{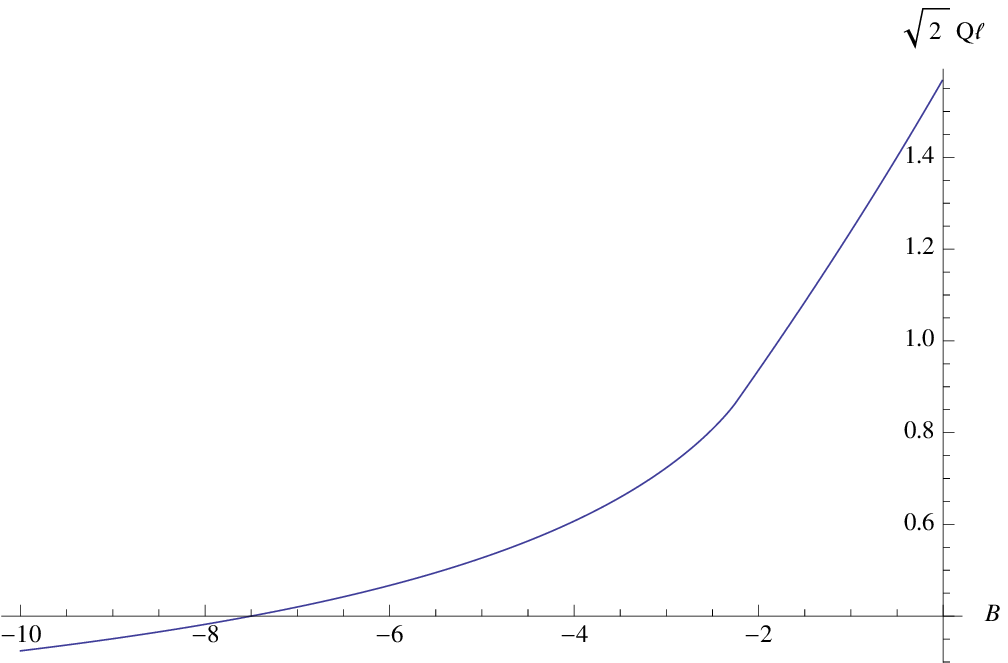}}
\noindent In summary, for negative $\lambda$ the connected solution exists for $\tilde{\ell}\in\left[0.4,\pi\right)$ whereas for non-negative $\lambda$, it exists for $\tilde{\ell}\in \left[\pi,3\pi\right)$.

We conclude this section by computing the difference between the entanglement entropy
of the connected solution and the disconnected one. The difference is
given by\foot{In the following we express the difference in the entropy in terms of $B$ instead of 
$\ell$ for reasons of convenience. Note that $B$ is a monotonic
function of $\ell$.}
\eqn\deltaS{\eqalign{& {32 Q^2\over L_x^2}\Delta S_{belt}\equiv {32 Q^2\over L_x^2}\left(S_{belt,\,con.}-S_{belt,\,disc.}\right)=\cr
&= \left({\lambda\over B}\right)^{3\over 2} \left\{ \left(-1+{2B\over 3}\right) +\right. \cr
&\left. + \int_0^1 dy \left(  {\left(9-2 B y^{{2\over 3}} \right)^2-
54 \left(-3+2 By^{{2\over 3}}\right)y^2 x'^2-9 \left(-9+8 B y^{{2\over 3}}\right)y^4 x'^4 \over 9y^2 \sqrt{1+y^2 x'^2} \left(9-2 B y^{{2\over 3}}+\left(9+4 B y^{2\over 3}\right)y^2 x'^2\right) }- {-9+2 B y^{{2\over 3}}\over 9 y^2}\right)\right\}
 } }
where we used \eqmgby\ and the fact the \slabactiongb\ is independent of $x$, to express the integral
in terms of $x'$ alone.
In Fig. 8 we plot $\Delta S$ for positive $B$ and in Fig. 9 for negative $B$. 
\ifig\loc{$\Delta S$ as a function of $B$ for $B\geq 0$.}
{\epsfxsize3.3in\epsfbox{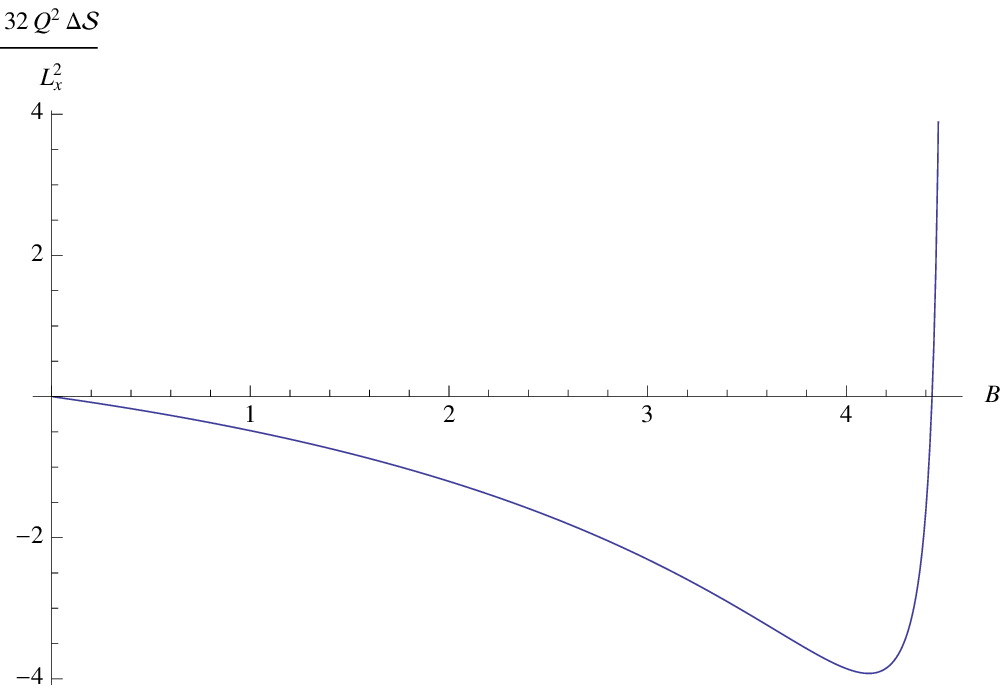}}
\ifig\loc{$\Delta S$ as a function of $B$ for $B<0$.}
{\epsfxsize3.3in\epsfbox{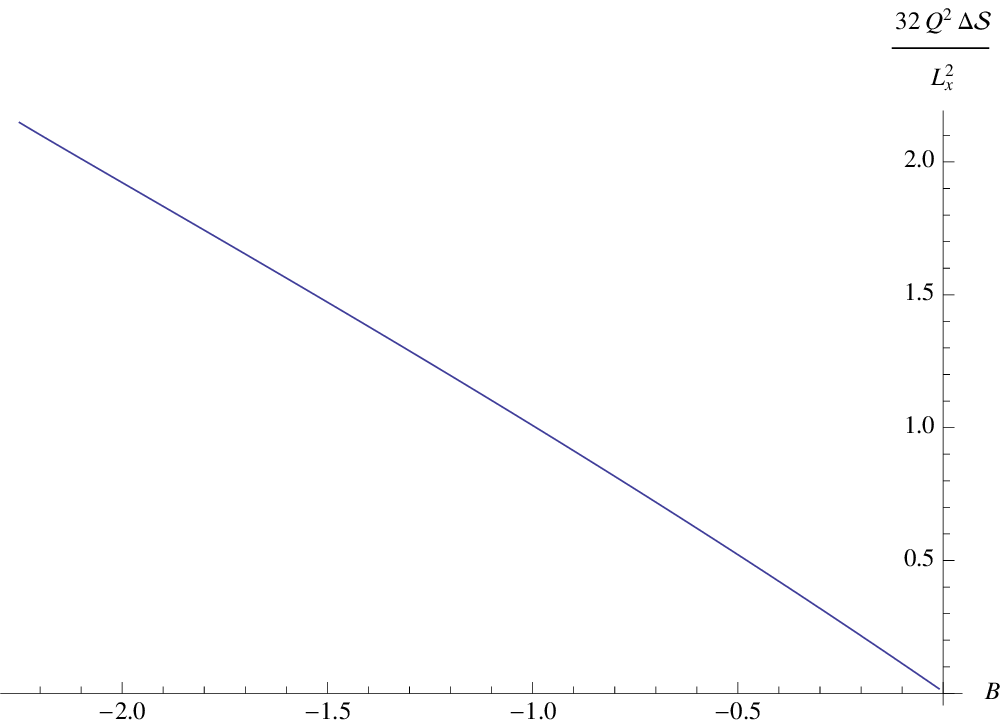}}
\noindent It is easy to see that for positive Gauss-Bonnet coupling, the connected solution--whenever available -- is favoured, while the contrary happens for negative values of the coupling.

\newsec{
%KS
Discussions and puzzles
%ENDKS
}

In this paper we computed holographic entanglement entropy in the background which corresponds to
the finite density state in the $\NN=4$ SYM theory with two equal $U(1)$ charges.
The computations were done for two simple geometries: a belt and a
sphere.
%KS
The puzzle we encounter is that 
we observed a phase transition in the case of a belt as a function of
its width, whereas the sphere does not. 
%It is always in the ``large size'' phase.

Such phase transitions in entanglement entropies 
%ENDKS 
have been observed before
in confining geometries \NishiokaGR\KlebanovWS\PakmanUI.
(Of course, to make intelligent statements about the phase transitions, one needs to define a UV-finite quantity;
one suitable definition in our case would involve subtracting the entanglement entropy in the vacuum, zero
density state.)
In \KlebanovWS\ a field theoretic explanation of this effect was proposed: the value of entanglement for the free glueball of mass $m$
is exponentially suppressed, $S\sim \exp(-m\ell)$; integrating this over the glue ball density of states with
Hagedorn behavior, $\rho(m)\sim \exp(\beta_H m)$ produces a phase transition at $\ell\sim \beta_H$.
The transition, which resembles Hagedorn phase transition, is between the $\OO(1)$ values of entanglement for sufficiently large 
values of $\ell$ and $\OO(N^2)$ values for small values of $\ell$, where the integral over glue ball states formally diverges.

%KS 
One would argue that the phase transition is a universal phenomenon
only depending on the size of the entangling surface compared to the
density. Indeed
%ENDKS.
in \PakmanUI\ a similar phase transition has been observed for the entanglement entropy of a sphere (disk)
in confining geometries dual to the 3+1 (2+1) dimensional theories.
This raises a %KS
second
%ENDKS
puzzle, since the (UV-finite part of) the free field entanglement entropy in this case is not exponentially suppressed,
but rather has an expansion in the inverse powers of the radius (see
e.g. \refs{\CasiniSR-\KlebanovYF}).

We do not have a resolution of this %KS
second
puzzle. 
Nor do we have a fundamental answer to the first. Instead, we add the 
observation that an understanding of these phase transitions
must be more complicated in that a third scale must play a role. 
%It is not as universal as the glueball Hagedorn transition interpretation would
%suggest. Here 
This third scale is supplied by the anisotropy of the belt. This must be so 
as one ought to be able to obtain the belt shape by deforming
the ball-entangling surface continuously into an ovoid. In
this process the phase-transition should suddenly
appear; qualitatively it must therefore depend on the eccentricity of
the ovoid. To check this explicitly is technically quite involved.  
One can, however, easily construct a different
geometric set-up which interpolates between the ball and the
belt and thereby argue that this interpretation is correct. 
This set-up is the annulus or two concentric balls. In the limit where the
inner radius vanishes one has the sphere, and in the double limit
where both radii are large, but the difference stays small, one has the
belt. It is well-known from the analogue study of holographic Wilson
loops that this annulus system has a phase transition between a
geometric embedding surface that is a half-torus for the belt-like
configuration and two nested sphere-like embeddings for the ball-like
configuration \refs{\GrossGK\ZaremboBU\OlesenJI\KimTD-\HirataJX}. This
suggests that the belt-like phase transition may be a consequence of the
third scale introduced by the anisotropy violation.
% more than a universal Hagedorn phenomenon.
% Namely, in the finite density state studied in this paper, the belt geometry gives rise to the phase transition,
% but the disk geometry does not!
%ENDKS
It would be interesting to construct a free-field model which would reproduce this behavior, although 
there is no guarantee such a free field description exists. 

Note, that for the case of states with Fermi surfaces, one expects the two geometries to produce the
same violation of the area law (and this is also the case in the holographic models of \HuijseEF). 
Of course, the area law is not violated in our case, so one can conclude that the ground state dual to the
geometry explored here is different from a Fermi liquid.

We would
 %KS
%also
finally
%ENDKS
 like to note that the addition of the Gauss-Bonnet term to the bulk Lagrangian
modifies the picture. 
In this case there are strong indications that the belt geometry can support connected surfaces for
arbitrarily large values of the distance between the
plates. Generally, adding %KS
higher derivative terms such as the Gauss-Bonnet term
amounts to 
%KS
%DROP modifying the UV structure of the theory, so one may argue that the pure Einstein-Hilbert
% gravity in the bulk gives rise to a non-generic infrared theory
the inclusion of $1/g^2_{YM}N$ and/or $1/N$ corrections in the dual
field theory. It suggests then that the phase-transition is a strong
coupling/large $N$ phenomenon which would be hard to capture in any
field theoretic approach.
The Gauss-Bonnet term is very special and we should be careful to
extrapolate generic lessons from it.
%ENDKS
%DROP On the other hand, adding higher derivative
%terms to the bulk lagrangian takes us out of the realm of established top-down AdS/CFT constructions.

\bigskip
\bigskip
{\noindent \bf Acknowledgements:} We would like to thank J. Jottar, I. Klebanov, D. Kutasov, Y. Liu, S. Sachdev, and J. Zaanen
for useful discussions.  M.K. and A.P. thank Aspen Center for Physics where part of this work was completed.
This work was supported in part by the NSF grant No. 1066293, a VIDI
grant from NWO and by the Dutch Foundation for Fundamental Research on Matter (FOM). The work of M.K. was partially supported by the ERC Advanced Grant "SyDuGraM", by IISN-Belgium (convention 4.4514.08) and by the ``Communaut\'e Fran\c{c}aise de Belgique" through the ARC program and the NSF grant No. 1066293.

\appendix{A}{Gauss-Bonnet, hyperscaling violation and entanglement entropy.}

\subsec{Solutions of GB gravity+matter which give rise to holographic hyperscaling violation.}

\noindent Here we show that geometries with hyperscaling violation are solutions of Gauss-Bonnet gravity with
matter and gauge fields. We consider the following five-dimensional action
\eqn\action{I_G= {1\over 16 \pi G_{N}} \int d^{4+1}x \sqrt{-g}\left[R+{12\over L^2}+{\lambda L^2\over 2}
\left(R_{\mu\nu\rho\sigma}^2-2 R_{\mu\nu}^2+R^2\right)-\LL^{m}\right]}
where $\lambda$ is the dimensionless Gauss-Bonnet coupling and $\LL^{m}$ is the matter Lagrangian
\eqn\LLmatter{\LL^{m}=Z(\Phi) F^2+{1\over 2} (\p \Phi)^2+V(\Phi)\,.}
Here $\LL(F^2)$ is an arbitrary function of the quadratic gauge invariant 
$F^2=F_{\mu\nu}F^{\mu\nu}$, with $F_{\mu\nu}=\p_\mu A_\nu-\p_\nu A_\mu$ 
the electromagnetic tensor field and $A_\mu$ the vector potential. $\Phi$ is a scalar field 
and $V(\Phi), \, Z(\Phi)$ are arbitrary functions. 
The action $I_G$ should in principle be supplemented with a boundary term to make the variational problem well-defined. For our purposes the detailed structure of the boundary term will not be necessary.

Varying the action with respect to the metric tensor $g_{\mu\nu}$ and the matter fields 
$\Phi, A_{\mu}$ yields the following equations of motion
\eqn\eqall{\eqalign{&G_{\mu\nu}-{6\over L^2}g_{\mu\nu}+{\lambda L^2\over 2} G_{\mu\nu}^{(2)}=T_{\mu\nu}^{m}\cr
&\nabla_{\nu}Z(\Phi)\left(F^{\mu\nu} \right)=0\cr
&\p_\mu\left(\sqrt{-g} g^{\mu\nu} \p_\nu\Phi\right)-\sqrt{-g}\left({\p V\over \p \Phi}+{\p Z\over\p \Phi} F_{\mu\nu}^2\right)=0
\, ,}}
where $G_{\mu\nu}\equiv  R_{\mu\nu}-{1\over 2} R g_{\mu}$ is the Einstein tensor, $T_{\mu\nu}^{m}$ is
the energy momentum tensor for the matter fields and $G_{\mu\nu}^{(2)}$ is the second order Lovelock tensor
\eqn\Lovelocktwo{\eqalign{G_{\mu\nu}^{(2)}&=2\left(R_{\mu\rho\kappa\lambda}R_{\nu}^{\,\rho\kappa\lambda}-2 R_{\mu\rho\nu\kappa}R^{\rho\kappa}-2 R_{\mu\rho}R^\kappa_{\, \nu}+R R_{\mu\nu}\right)-\cr
&-{1\over 2} \left(R_{\kappa\lambda\rho\sigma}^2-2 R_{\kappa\lambda}^2+R^2\right)g_{\mu\nu}
\,.}}
The matter energy momentum tensor has the form
\eqn\matterem{T_{\mu\nu}^{m}=-{1\over 2}g_{\mu\nu}\LL^{m} +2(\p_\mu\Phi)(\p_\nu\Phi)+2 Z(\Phi) F_{\mu}^{\,\rho}F_{\nu\rho} \,.}
We search for solutions of the form
\eqn\metgena{\eqalign{ds^2&=e^{2 A(r)} \left(dr^2 - e^{2 B(r)} dt^2 + dx_i^2\right) \cr
A_t&=A_t(r) \quad A_i=0,\, i=1,2,3,\quad \Phi=\Phi(r)
\,. }}
Substituting the ansatz \metgena\ into the equations of motion \eqall\ yields
\eqn\eqmansatz{\eqalign{T_t^{t,\,m}-T_i^{i,\, m}&=2 F_{tr} F^{tr}Z(\Phi)=-e^{-2 A(r)} \left(3 A'(r)B'(r)+B'(r)^2+B''(r)\right)+\cr
&+2 \lambda L^2 e^{-4 A(r)} A'(r)\left[B'(r)\left(A'(r)^2+A'(r)B'(r)+2A''(r)\right)+A'(r)B''(r)\right]\cr
T_r^{r,\,m}-T_t^{t,\, m}&=e^{-2 A(r)}(\p_r\Phi)^2 =\cr
&=3 e^{-2 A(r)}\left(1-2\lambda L^2 e^{-2A(r)}A'(r)^2\right)\left(A'(r)^2+A'(r)B'(r)-A''(r)\right)\cr
&\p_r \left(e^{5 A+B}Z(\Phi) F^{tr} \right)=0\cr
e^{-5 A(r)-B(r)}&\p_r\left(e^{3 A(r)+B(r)}\p_r\Phi\right)={\p V\over\p\Phi}+2{\p Z\over\p \Phi} F_{tr}F^{tr}
\,.}}
%The number in front of the last term in the last line of the equation above should be checked.
We can easily solve the Gauss Law constraint for $Z(\Phi)F^{tr}$ 
in the third line of \eqmansatz\ to get
\eqn\ZFsol{ Z(\Phi)F^{tr}=Q e^{-5 A(r)-B(r)}\,} 
where $Q$ is the total charge of the solution. Substituting the result into the first line
of \eqmansatz\ determines $F_{tr}$ in terms of $A(r),\, B(r)$ and their first and second derivatives.
To be specific
\eqn\Ftr{\eqalign{F_{tr}&=-{e^{3 A(r)+B(r)}\over 2 Q} \left(3 A'(r)B'(r)+B'(r)^2+B''(r)\right)+\cr
&+2 \lambda L^2 {e^{ A(r)+B(r)}\over 2 Q} A'\left[B'\left(A'^2+A'B'+2A''\right)+A'(r)B''(r)\right] 
\,}}
We can then use the solution \Ftr\ to solve for $Z(\Phi(r))$ from \ZFsol,
\eqn\Zsol{Z(\Phi)= {2 Q^2 e^{-4A(r)} \over \left(3 A'B'+B'^2+B''\right)-
2 \lambda L^2 e^{ -2A(r)} A'(r)\left[B'(r)\left(A'(r)^2+A'(r)B'(r)+2A''(r)\right)+A'B''\right] } \,.}
In addition, note that the second line of \eqmansatz\ gives a solution for $\p_r\Phi$ in terms of the metric 
functions $A(r)\, B(r)$, \ie, 
\eqn\phip{\Phi'(r)=\sqrt{3\left(1-2\lambda L^2 e^{-2A(r)}A'(r)^2\right)\left(A'(r)^2+A'(r)B'(r)-A''(r)\right)}\,.}
Eq. \phip\ can in principle be integrated to give an expression for $\Phi(r)$ in
terms of the radial variable. Then it can be inverted to express $r$ in terms of $\Phi$ and
determine $Z(\Phi)$.
Finally, substituting \phip\ together with \Ftr\ and \Zsol\ into the last line of \eqmansatz\
an expression for $\p V\over\p r$ in terms of $A(r),\, B(r)$ can be easily determined. Simply rewrite 
the derivatives in terms of $\Phi$ in \eqmansatz\ as ${\p\over\p\Phi}={1\over \Phi'(r)}{\p\over\p r}$
and solve for ${\p V\over\p r}$ in \eqmansatz.

It is clearly not very difficult to produce solutions of the type \metgena\ from the action \action.
Not all of these solutions however are necessarily consistent either as gravitational solutions or as 
dual descitpions of quantum field theories. Absence of singularities, appropriate fall-off conditions 
for the fields, etc. restrict the set of consistent solutions.
One of the simplest consistency checks, is the null energy condition  
\eqn\nec{T_{\mu\nu}^m u^\mu u^\nu\geq 0\, ,}
where $u^\mu$ is an arbitrary null vector of the spacetime \metgena, \ie, $g_{\mu\nu} u^\mu u^\nu=0$.
For the metric in \metgena\ the null energy condition essentially amounts to requiring the first
and second line of \eqmansatz\ to be non-negative
\eqn\necgen{T_t^{t,\,m}-T_i^{i,\, m}\leq 0\qquad T_r^{r,\,m}-T_t^{t,\, m}\geq 0\,.}

We will now focus on a specific choice of metric which belongs to the class of metrics 
described by \metgena\ and violates hyperscaling (for more details on these solutions and their condensed matter applications see \refs{\HuijseEF\OgawaBZ\ShaghoulianAA\DongSE\NarayanHK\KimNB\HashimotoTI-\PerlmutterHE})
\eqn\methyper{ds^2= r^{-2 (3-\theta)/3} \left( dr^2-r^{-2 (z-1)} dt^2 + dx_i^2\right) \, .}
The starting point will be to analyze the null energy condition \necgen. Substituting
the expressions for $A(r),\, B(r)$ from \methyper\ in \necgen\ yields
\eqn\nechyper{\eqalign{{(\theta-3)\over 3}r^{-2\theta\over 3}\left(3+\theta-3z\right)\left(1-{2\over 9}\lambda L^2 (\theta-3)^2 r^{-2\theta\over 3}\right)&\geq 0\cr
(z-1)(3-\theta+z)r^{-2\theta\over 3}\left(1+{2\over 27}\lambda L^2 r^{-2\theta\over 3} {(\theta-3)^2(\theta-6-3 z)\over 3-\theta+z}\right)&\geq 0
\, .}}
To determine the allowed range of values for the parameters $(\theta,\,z)$ from \nechyper\ we make the following assumptions:
\item  1 that \methyper\ is meaningful for all $r$, ranging from zero to very large values and
\item  2 that the theory is also consistent at the $\lambda= 0$ point.

\noindent Eq. \methyper\ then reduces to 
\eqn\necbounds{\eqalign{ (\theta-3)(3-3z+\theta)\geq 0\qquad\qquad\qquad\qquad\qquad\lambda&\leq 0\cr
(z-1)(3-\theta+z)\geq 0\qquad\qquad\qquad {\theta-6-3 z\over 3-\theta+z}&\leq 0
}}
The left column of \necbounds\ provides the constraints on $\theta,\, z$ obtained
form the null energy condition in the case of Einstein-Hilbert gravity. It is easy to see that
the right column does not yield additional constraints on $\theta,\,z$. It only restricts the value
of the Gauus-Bonnet $\lambda$ coupling to be negative.

It is clear that the geometry of \adsconfhyper\ produced by taking the double-scaling limit 
\doublescalinglimit\ remains a solution of Gauss-Bonnet gravity. The null energy condition in this case
yields $\lambda L^2\leq {9\over 2}$.

\subsec{Holographic entanglement entropy and hyperscaling violation in GB gravity}

\noindent In the following we compute holographic entanglement entropy
in the GB gravity.
We restrict to $3+1$ dimensional boundary (so that the GB term does not vanish in the bulk).
We will consider the general form of the metric \metgena.
The example of a metric that violates hyper scaling is given by \methyper.
%\eqn\hypervio{  ds^2 = r^{-2 (3-\theta)/3} \left( -r^{-2 (z-1)} dt^2+ dr^2 + dx_i^2\right)  }
The holographic entanglement entropy is equal to the extremal value of the
action
\eqn\acsig{  S= \int d^3x \sqrt{\det G_\Sigma } \left(1+ \lambda R_\Sigma \right) }
where $G_\Sigma$ is the induced metric.

We consider the case of a slab as defined in section 2, with induced metric
\eqn\gsigma{  ds_\Sigma^2=  r^{-2 (3-\theta)/3} \left( (1+ (x')^2 ) dr^2 +dx_2^2+dx_3^2  \right)}
where we defined $x\equiv x_1$.
In this case the action \acsig\ reduces to 
\eqn\acslab{ S= \int dr e^{3 A(r)} \sqrt{ (1+(x')^2)}
   \left(1{-}\frac{2 \lambda e^{-2 A(r)} [A'(r)^2 (1+(x')^2){+}2 (1+(x')^2) A''(r){-}A'(r) (x'^2)' ]}{(1+(x')^2)^2}\right)
   }
One can solve for $x'$ and expand near the boundary $x=0$
\eqn\xpexp{  x'^2 = \frac{81 r_0^{2\theta-6} r^{6-\frac{2 \theta}{3}} \left(1+\frac{9 r^{2 \theta/3}}{\lambda (\theta-3)^2}\right) 
 }{4 \lambda^2 (\theta-3)^4} +\ldots   }
For $0\leq \theta\leq 3$,  $x'^2$ does not contribute to the UV (small $r$) divergence of \acslab.

The most interesting divergence in \acslab\ comes from the $e^{3 A(r)}$
term and gives rise to the $S\simeq (\epsilon/\ell)^{\theta-2}$ behavior, similarly to 
the Einstein-Hilbert case.
The case $\theta=d-1=2$ gives rise to a logarithmic term, $S\simeq \log\ell/\epsilon$,
just as it did in the Einstein-Hilbert ($\lambda=0$) case.
Interestingly, in the Gauss-Bonnet case there are terms proportional to $\lambda$ which exhibit stronger divergence,
$S\simeq (\epsilon/\ell)^{{\theta\over 3}-2}$.
Such terms, however, do not produce logarithmic violations of the entanglement entropy. Moreover,
they are expected to vanish when the contribution from the boundary term neglected in \acsig\ is included.

%Kuritsis etal, perhaps embed \bhmetric\ in Gauss-Bonnet gravity? What would be the role of the additional parameter?

\footatend\vfill\supereject\immediate\closeout\rfile\writestoppt
\baselineskip=14pt\centerline{{\bf References}}\bigskip{\frenchspacing%
\parindent=20pt\escapechar=` \input refs.tmp\vfill\eject}\nonfrenchspacing

\bye